\documentclass[openacc]{rsproca_new}

\usepackage{graphicx}
\usepackage{subcaption}
\usepackage{amsmath}
\usepackage{amssymb}

\begin{document}

%%%% Article title to be placed here
\title{Classically Bound and Quantum Quasi-Bound States of an Electron on a Plane Adjacent to a Magnetic Monopole}

\author{%%%% Author details
J. Martin\(^1\), A. Baskerville\(^{1,2}\), V. L. Campo\(^3\), J. Minns\(^1\), J. Pooley\(^1\), S. T. Carr\(^1\),  C. A. Hooley\(^4\), G. M\"oller\(^1\), J. Quintanilla\(^1\)}

%%%%%%%%% Insert author address here
\address{$^{1}$Physics of Quantum and Materials Group, School of Engineering, Mathematics and Physics, University of Kent, Canterbury, Kent CT2 7NH, United Kingdom \\
$^{2}$Kvantify, 1 Long Ln, London SE1 4PG, United Kingdom\\
$^{3}$Departamento de F\'isica, Universidade Federal de S\(\tilde{\text{a}}\)o Carlos, Rodovia Washington Luiz, km 235, Caixa Postal 676, 13565-905, S\(\tilde{\text{a}}\)o Carlos, S\(\tilde{\text{a}}\)o Paulo, Brazil\\
$^{4}$Centre for Fluid and Complex Systems,
Coventry University, Coventry CV1 2TT, United Kingdom}

%%%% Subject entries to be placed here %%%%
\subject{quantum physics, mathematical physics}

%%%% Keyword entries to be placed here %%%%

%%%% Insert corresponding author and its email address}
\corres{S. T. Carr\\
\email{s.t.carr@kent.ac.uk}}

%%%% Abstract text to be placed here %%%%%%%%%%%%
\begin{abstract}
In three-dimensional space an electron moving in the field of a magnetic monopole has no bound states.  In this paper we explore the physics when the electron is restricted to a two-dimensional plane adjacent to a magnetic monopole.  We find bound states in the classical version of the problem and quasi-bound states in the quantum one, in addition to a continuum of scattering states.  We calculate the lifetimes of the quasi-bound states using several complementary approximate methods, which agree well. The threshold monopole magnetic charge required to realise a single quasi-bound state is approximately \(18Q_D\), where \(Q_D\) is the magnetic charge of a Dirac monopole.  We examine the feasibility of achieving this magnetic charge in currently available monopole analogues:\ spin ice, artificial spin ice, magnetic needles, image charges in magnetoelectric materials, and emergent quantum excitations in Josephson junction arrays or superconducting films. 
\end{abstract}

%%%%%%%%%% Insert the texts which can accommodate on first page in the tag "fmtext" %%%%%

%\begin{fmtext}
%\end{fmtext}

%%%%%%%%%%%%%%% End of first page %%%%%%%%%%%%%%%%%%%%%

\maketitle %leave this command here for royal society style

%########## Introduction section ##########
\section{Introduction} 
In 1931 Dirac imagined a magnetic monopole as a defect by constructing a vector potential that led to a monopolar field everywhere in space but which was singular on a single line, an infinitely thin undetectable solenoid commonly referred to as a Dirac string \cite{RefWorks:33}. These Dirac monopoles have yet to be found, if they can be found at all.

Although no true Dirac monopoles have been found thus far, there are several monopole-like sources, which resemble true monopoles in many regards:
\begin{itemize}
    \item Spin ice materials.  These consist of magnetic moments (spins) arranged on a pyrochlore lattice.  Flipping a spin in the ground state can create a pair of magnetic monopoles on the two tetrahedra adjacent to the spin; due to frustration, these monopoles are deconfined.  Subsequent spin flips can either move them away from one another or cause adjacent monopoles of opposite magnetic charges to recombine \cite{Castelnovo2008,Bramwell2020}.
    \item Artificial spin ice.  This is constructed using nanoscale dipole magnets arranged on the links of a lattice, with the two poles of each magnet sitting on neighbouring lattice sites.  Flipping the polarity of one of the magnets produces an overall magnetic charge on the lattice site at one of its ends and an equal and opposite magnetic charge at the site on the other.  At each of these lattice sites it now appears that a magnetic monopole is present; as with spin ice, flipping the polarities of magnets adjacent to these sites can cause the monopoles to move away from one another \cite{Moller2006,Ladak2010,Chern2011,Wang2006}.
    \item Magnetic needles. Another way to create an artificial magnetic monopole is via a nanoscale ferromagnetic needle. As the ends of the magnetic needle are far apart, the field produced near one end of the needle is very similar to that of a magnetic monopole \cite{RefWorks:17}.
    \item Image charges in magnetoelectric materials.  In this case the effective magnetic monopole is made by exploiting the parallel linear coupling (${\bf E} \cdot {\bf B}$ coupling) that occurs between the electric and magnetic fields in certain materials, including some topological insulators.  As a result of this, an electric charge placed above the material's surface induces a magnetic monopole at the image charge location inside the material \cite{Khomskii2014,Fechner2014,Meier2019}.
    \item Emergent quantum excitations in Josephson junction arrays and superconducting films.  It was pointed out recently \cite{Trugenberger2020} that magnetic monopoles also arise in the form of quantum instantons in Josephson junction arrays, and that the phenomenon of superinsulation could be viewed as the condensation of such monopoles \cite{Diamantini2021}.
\end{itemize}
It should be noted that, in all but the last of these examples, the effective magnetic monopoles do not typically satisfy the Dirac quantisation condition.

It is interesting to ask whether magnetic monopoles could be detected through their effect on the motion of nearby charged particles, e.g.\ electrons.  In 1896 Birkeland carried out an experiment of this sort, observing that when the pole of an electromagnet is placed at the opposite end of a  Crookes tube from the cathode the trajectories of the cathode rays converge on the magnetic pole \cite{Birkeland1896}.  Poincar{\'e} later noted \cite{Poincare1896} that the motion of the electrons in the Crookes tube is as if a magnetic monopole were present.  He showed that, in the presence of the field of a magnetic monopole, electrons follow geodesics lying on the surface of a cone.  It follows that there are no closed orbits for an electron moving in three dimensions in the field of a magnetic monopole.

In this paper we explore whether this lack of bound states can be remedied by restricting the electron's motion to a plane, e.g.\ using the kind of two-dimensional electron gas system routinely used for quantum Hall effect measurements \cite{Ando1975,Leblanc2012}. This problem is usually studied in the presence of a uniform perpendicular applied magnetic field:\ in the classical analysis of such a system all states are bound, and this remains true when the electron's motion is treated quantum mechanically.  But what if we were to replace the uniform magnetic field with the field of a magnetic monopole located below the plane?  Would all classical orbits in this case be circular?  Would the quantum version of the problem yield bound states, quasi-bound states, or no bound states at all?  If there are quasi-bound states, what are their lifetimes?  We should note that a related problem of the potential superconducting pairing of a pair of electrons confined in planes on either side of an emergent magnetic monopole has been considered recently \cite{Diamantini2022}.  We will discuss the relationship of our work to this previous study in section \ref{sec:discussion} after we have presented our results.

%We should note that a related problem was already considered in ref.~\cite{Diamantini2022}.  In that work the authors were concerned with the potential superconducting pairing of a pair of electrons confined in planes on either side of an emergent magnetic monopole.  Their work differs from ours in three important respects:\ first, they include in their treatment the Zeeman coupling of the electrons' spins to the monopole's magnetic field; second, they solve only for the case where the centre-of-mass co-ordinate ${\bf R}$ is set to ${\bf 0}$; and third, they neglect the ${\bf A}^2$ term in the kinetic part of the Hamiltonian, on grounds that do not apply in the case that we consider here.

The remainder of this paper is structured as follows.  In section 2 we present the model and introduce our notation.  In section 3 we obtain the classical orbits using the Euler-Lagrange method, and explore their semi-classical quantisation in the Bohr-Sommerfeld scheme.  In section 4 we address the quantum version of the problem, using both the WKB method and a numerical approach.  In section 5 we estimate the lifetimes of the quasi-bound states in the quantum problem using various complementary techniques.  In section 6 we discuss our results, and in section 7 we
conclude.

%######## end of introduction section#######

%########Main body of text###############
\section{Preliminaries}
\begin{figure}
    \centering
    \includegraphics[width=0.8\textwidth]{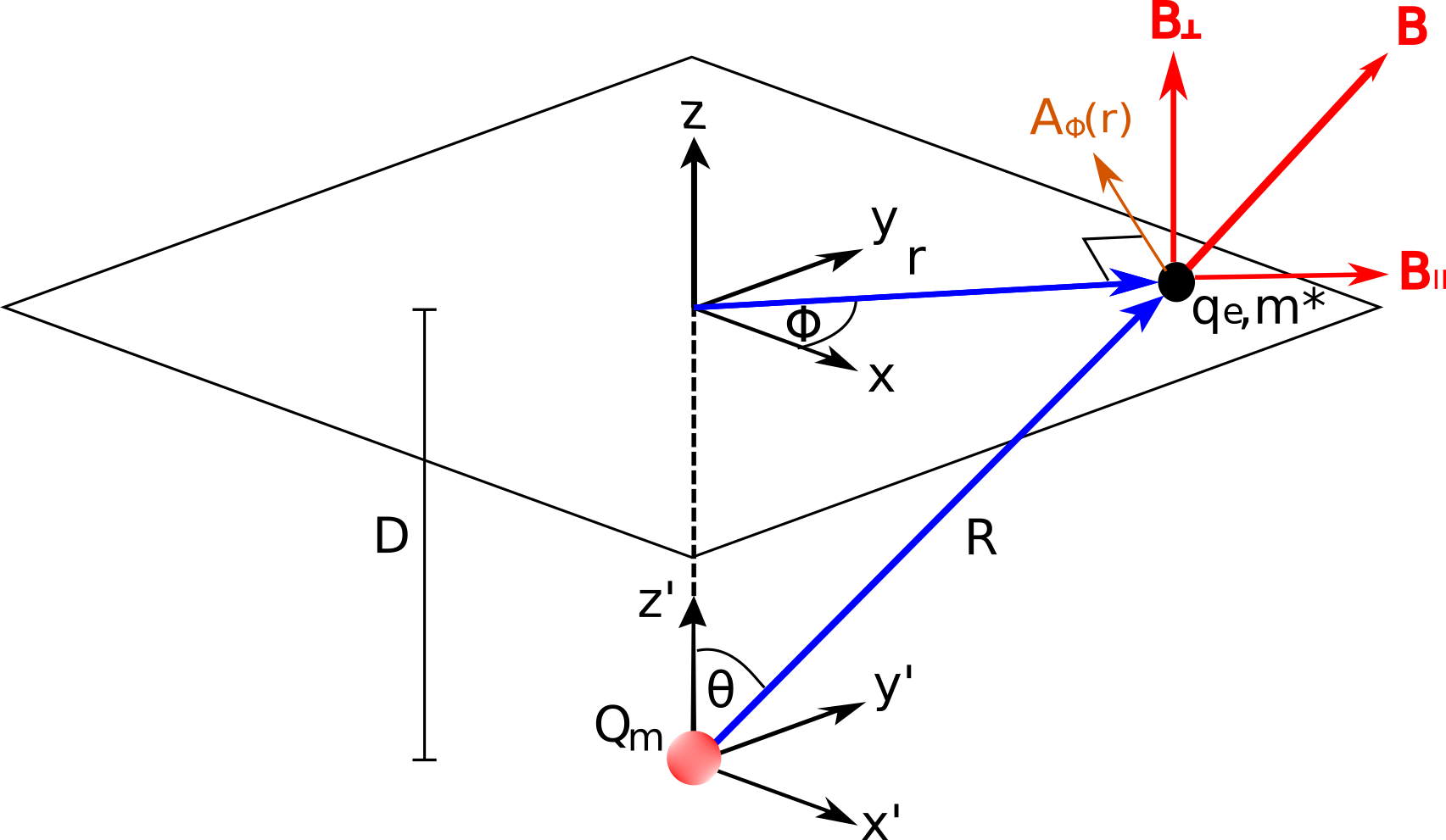}
\caption{An electron, of electric charge $q_e = -\vert q_e \vert$ and mass $m^{*}$, moving in a plane a distance $D$ above a magnetic monopole of charge $Q_{m}$.  We use cylindrical coordinates with the axis passing through the monopole and the vertical distance measured from the plane ($z$) or alternatively from the monopole ($z'$). We decompose the magnetic field \(\mathbf{B}\) into in-plane ${\bf B}_\parallel$ and out-of-plane \({\bf B}_\perp\) components. The vector potential corresponding to the perpendicular component is denoted by \(\mathbf{A}=A_\phi(r)\hat{\mbox{\boldmath $\phi$}}\), where \(\hat{\mbox{\boldmath $\phi$}}\) is the unit vector in the direction of increasing azimuthal coordinate.}
\label{fig graphical representation of the system}
\end{figure}
We consider an electron of effective mass \(m^*\) and electric charge $q_e = - \vert q_e \vert$ confined to a plane situated a distance \(D\) above a magnetic monopole of charge \(Q_m\), as shown in Fig.~\ref{fig graphical representation of the system}.  In this context it will not matter to us whether or not the magnetic monopole satisfies the Dirac quantisation condition, since the electron is sensitive only to the magnetic field profile in the plane where it is confined:\ if this is magnetic-monopole-like, that will be sufficient.

Gauss's law for magnetism (the no-monopole law) can be rewritten to include magnetic monopoles \cite{RefWorks:36,RefWorks:37},
\begin{equation}
    \oint\limits_{\cal S}\mathbf{B}.d\mathbf{S}= \mu_0 Q_m,
    \label{Eq maxwell monopole}
\end{equation}
 where ${\cal S}$ is any closed surface, $Q_m$ is the net magnetic charge enclosed within ${\cal S}$, and \(\mu_0\) is the permeability of free space. The magnetic field produced by the monopole is spherically symmetric; the magnetic field for a single magnetic monopole at the origin is thus
\begin{equation} 
\mathbf{B}=\frac{\mu_0 Q_m}{4 \pi R^2}\hat{{\bf R}},  
\label{eq monopole magnetic field}
\end{equation}
where \(\hat{{\bf R}}\) is the unit vector in the (three-dimensional) radial direction.
If we neglect the Zeeman effect, as we do throughout this paper, the only part of the magnetic field that affects the motion of the electron is the component perpendicular to the plane, \({\bf B}_\perp\):
\begin{equation} 
 \mathbf{B}_\perp = \left|\mathbf{B}\right|\frac{D}{\sqrt{r^2+D^2}} \hat{{\bf z}} = \frac{\mu_0 Q_m}{4 \pi} \frac{D}{\left( r^2 + D^2 \right)^{3/2}} \hat{{\bf z}}.
 \label{eq mag field perp}
\end{equation}
We therefore write the vector potential for this component of the field alone, i.e.\ ${\bf B}_\perp = \nabla \times {\bf A}$.  Choosing the Coulomb gauge, where \(\mathbf{\nabla} \cdot \mathbf{A}=0\), we may write the vector potential as a purely azimuthal field \(\mathbf{A}=A_\phi(r)\hat{\mbox{\boldmath $\phi$}}\).  It is given by
\begin{equation}
    A_\phi(r) = \frac{\mu_0 Q_m}{4 \pi r}\left[1-\frac{D}{\sqrt{r^2+D^2}}\right];
    \label{Eq A dimesionfull}
\end{equation}
the derivation of this equation is given in Appendix \ref{Appendix vector potential}.  The Hamiltonian \cite{RefWorks:1,RefWorks:29,RefWorks:34,RefWorks:35} governing the electron's motion is thus
\begin{equation}
        H_{\rm 2D} = \frac{1}{2m^{*}}\left(p_r^2 +\left[\frac{1}{r}p_\phi -q_eA_\phi(r)\right]^2\right),
    \label{Eq hamiltonian}
\end{equation}
where $p_r$ and $p_\phi$ are respectively the radial and azimuthal components of the electron's canonical momentum.  It follows from the Hamiltonian's lack of explicit dependence on $\phi$ that the associated canonical momentum $p_\phi$ is a conserved quantity.  We shall use the notation $p_\phi = \hbar M$, even in the classical case where $p_\phi$ is not quantised --- in that case $M$ will serve simply as a dimensionless measure of $p_\phi$. 

This problem contains a natural unit of length, $D$, and thus a natural unit of energy,
\begin{equation}
E_0 = \frac{\hbar^2}{2 m^{*} D^2}, \label{energyscale}
\end{equation}
and also of time,
\begin{equation}
t_0 = \frac{\hbar}{E_0}.
\label{eq:t0}
\end{equation}
We also introduce the dimensionless parameter $\lambda$, which measures the strength of the magnetic monopole in units of twice the Dirac monopole charge, i.e.\ $Q_m = 2 \lambda Q_D$, where
\begin{equation}
Q_D = \frac{2 \pi \hbar}{\mu_0 \vert q_e \vert}. \label{eq dirac charge}
\end{equation}
We define dimensionless measures of distance, $\rho$, and energy, $\epsilon$, by the following equations:
\begin{eqnarray}
r & = & \rho D; \label{r_scale}\\
E & = & \epsilon E_0. \label{E_scale}
\end{eqnarray}
When we talk about lifetimes of quasi-bound states, we will also measure time in dimensionless units (units of $t_0$) until section \ref{sec:discussion} when we talk about potential realisations.

\section{Classical and semi-classical solutions}

\subsection{Classical solution}
Hamilton's equations for the Hamiltonian $H_{\rm 2D}$ are:
\begin{eqnarray}
& \displaystyle {\dot r} = \frac{\partial H_{\rm 2D}}{\partial p_r} =
\frac{p_r}{m^{*}}; & \label{hameq1} \\
& \displaystyle {\dot \phi} = \frac{\partial H_{\rm 2D}}{\partial p_\phi} =
\frac{1}{m^{*} r^2} \left( p_\phi - q_e r A_\phi \right); & \label{hameq2} \\
& \displaystyle {\dot p_r} = - \frac{\partial H_{\rm 2D}}{\partial r} = - \frac{1}{2 m^{*}} \frac{\partial}{\partial r} \left[ \left( \frac{p_\phi}{r} - q_e A_\phi \right)^2 \right], \label{hameq3}
\end{eqnarray}
where a dot represents a (total) time-derivative.  (As noted above, ${\dot p}_\phi = 0$.)  Using (\ref{hameq1}) to eliminate $p_r$ from (\ref{hameq3}), we obtain the radial equation of motion
\begin{equation}
    m^*\ddot{r}=-\frac{1}{2m^*}\frac{\partial}{\partial r} \left[ \left(\frac{p_\phi}{r}-q_eA_\phi\right)^2 \right].
    \label{Eq equation of motion}
\end{equation}
By comparison with the general expression
\begin{equation}
m^{*} {\ddot r} = - \frac{\partial {\tilde V}_{\rm cl}}{\partial r},
\end{equation}
we find that the effective one-dimensional potential energy, \({\tilde V}_\text{cl}(r)\), is given by
\begin{equation}
    \centering
    {\tilde V}_\text{cl}(r)=\frac{1}{2m^*}\left(\frac{p_\phi}{r}-q_eA_\phi(r)\right)^2.
    \label{Eq classical potential SI units}
\end{equation}

Using the form of \(A_\phi(r)\) given in (\ref{Eq A dimesionfull}) and the dimensionless quantities introduced above, we find that the potential energy in units of $E_0$ is given by the dimensionless function
\begin{equation}%scaled potential
    \centering
    V_\text{cl}(\rho) \equiv \frac{{\tilde V}_{\rm cl}}{E_0} = \frac{\lambda^2}{\rho^2}\left[\frac{M}{\lambda}+\left(1-\frac{1}{\sqrt{1+\rho^2}}\right)\right]^2.
    \label{eq classical scaled potential}
\end{equation}
We recall that $p_\phi = \hbar M$ and $q_e = -\vert q_e \vert$; the latter of these facts accounts for the apparent change from a difference in (\ref{Eq classical potential SI units}) to a sum in (\ref{eq classical scaled potential}).

\begin{figure}
    \begin{subfigure}{0.49\linewidth}
    \includegraphics[width=\linewidth]{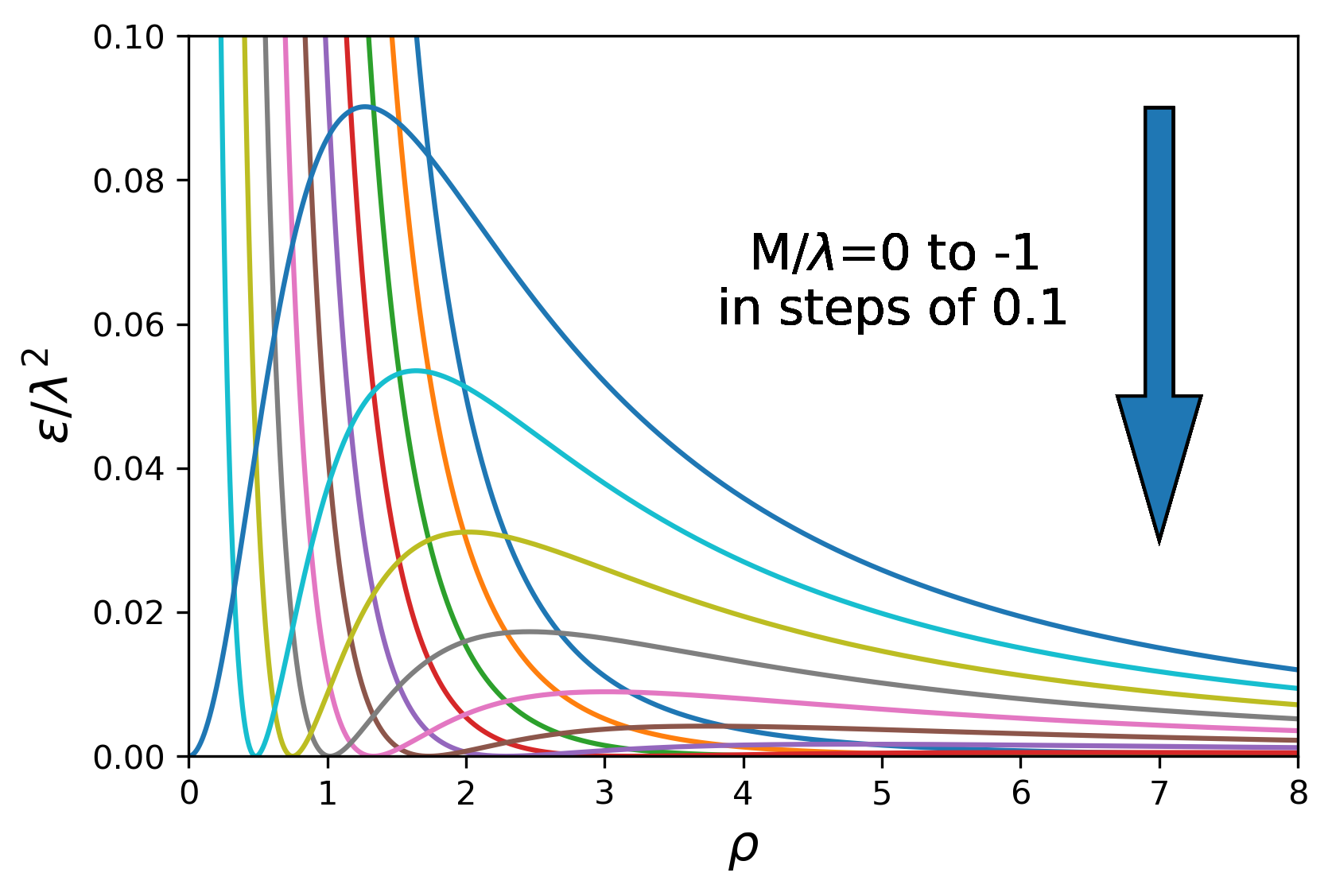}
    \caption{}
    \end{subfigure}
    \begin{subfigure}{0.49\linewidth}
    \includegraphics[width=\linewidth]{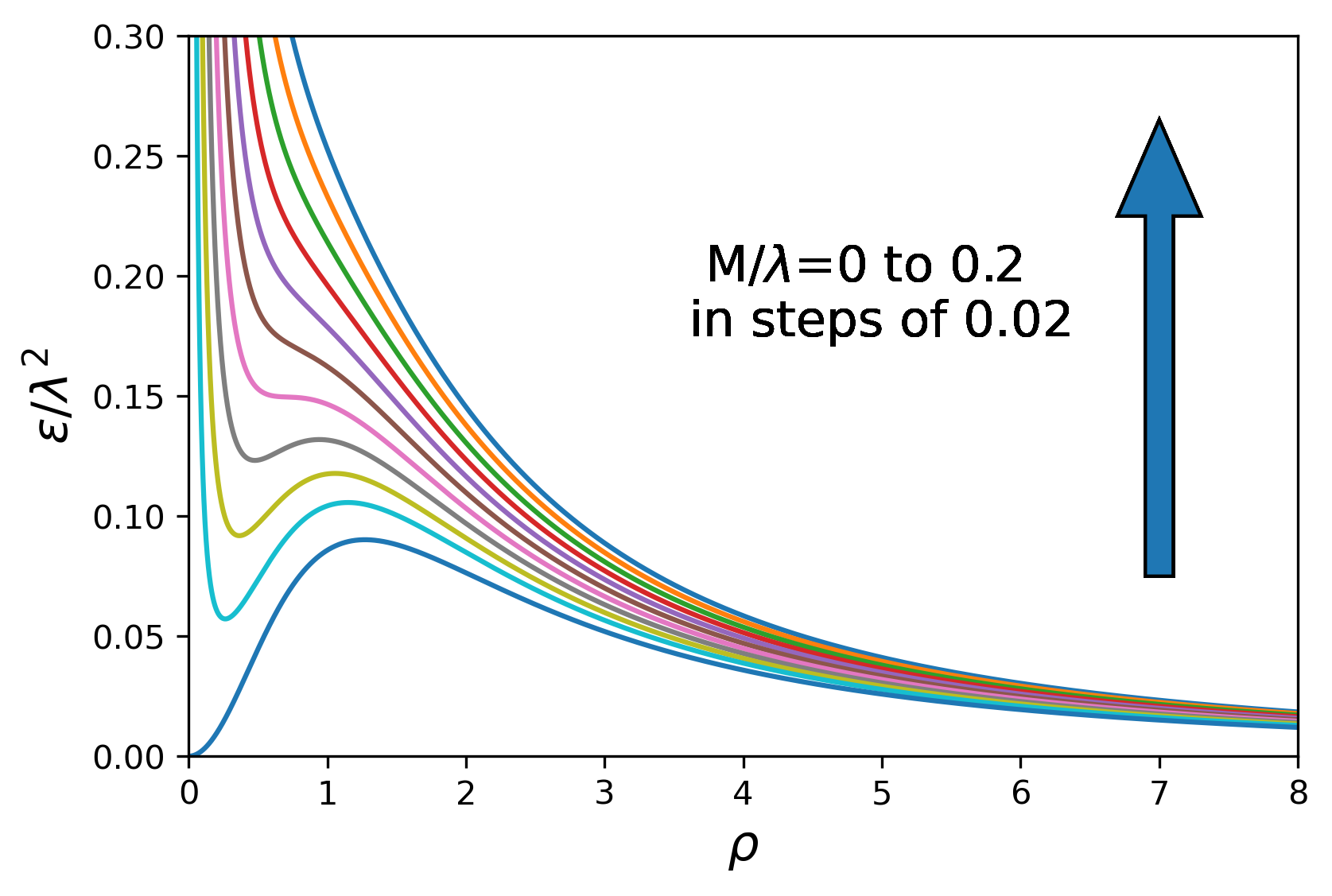}
    \caption{}
    \end{subfigure}
    \caption{The effective classical potential governing the radial motion of the electron, \(V_\text{cl}(\rho)/\lambda^2\).  The shape of this potential depends only on the ratio $M/\lambda$, where $\hbar M$ is the canonical angular momentum of the electron and $\lambda$ is the strength of the magnetic monopole in units of twice the Dirac monopole charge.  (a) \(V_\text{cl}(\rho)/\lambda^2\) for \(M/\lambda\) decreasing from 0 to $-1$ in steps of 0.1 in the direction of the arrow. (b) \(V_\text{cl}(\rho)/\lambda^2\) for \(M/\lambda\) increasing from 0 to 0.2 in steps of 0.02 in the direction of the arrow.}
    \label{Fig Classical Potentials}
\end{figure}

The shape of this effective potential depends only on the ratio $M/\lambda$.  For negative $M/\lambda$, the potential has a zero at finite radius $\rho$, while for positive $M/\lambda$ it does not.  Examples of the effective potential for negative and positive values of $M/\lambda$ are shown in Figs.~\ref{Fig Classical Potentials}(a) and \ref{Fig Classical Potentials}(b) respectively.  In both cases, we see the formation of a local minimum at finite $\rho$; in the negative-$M/\lambda$ case this persists to arbitrarily large values of $\vert M/\lambda \vert$, whereas in the positive-$M/\lambda$ case it persists only up to
\begin{equation}
\left( \frac{M}{\lambda} \right)_{\rm max} = \left[ 2 \left( \frac{2}{3} \right)^{3/2} - 1 \right] \approx 0.089.
\end{equation}

\begin{figure}
\centering
\begin{subfigure}{0.51\linewidth}
\includegraphics[width=\linewidth]{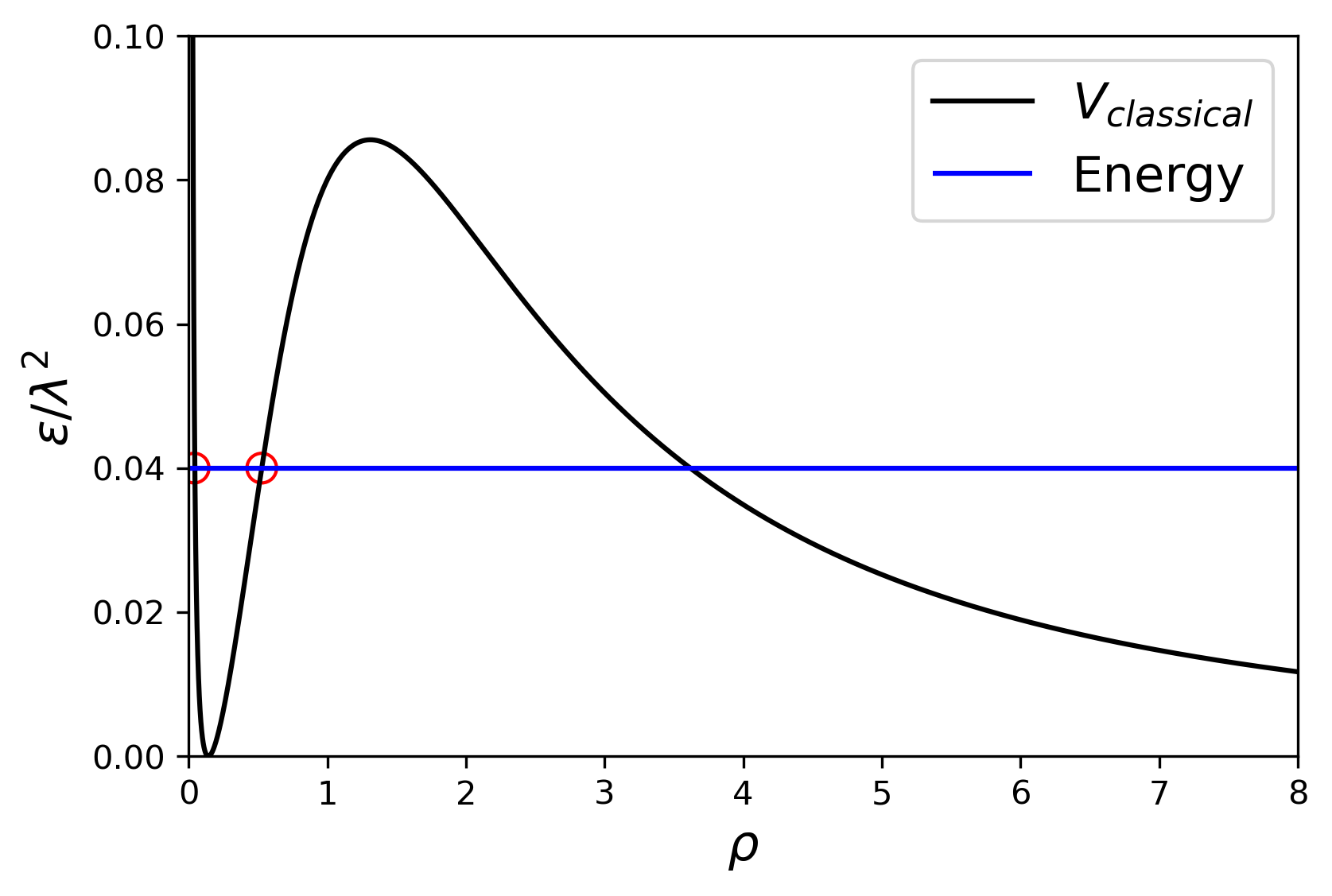}
\caption{}
\end{subfigure}
\begin{subfigure}{0.49\linewidth}
\includegraphics[width=\linewidth]{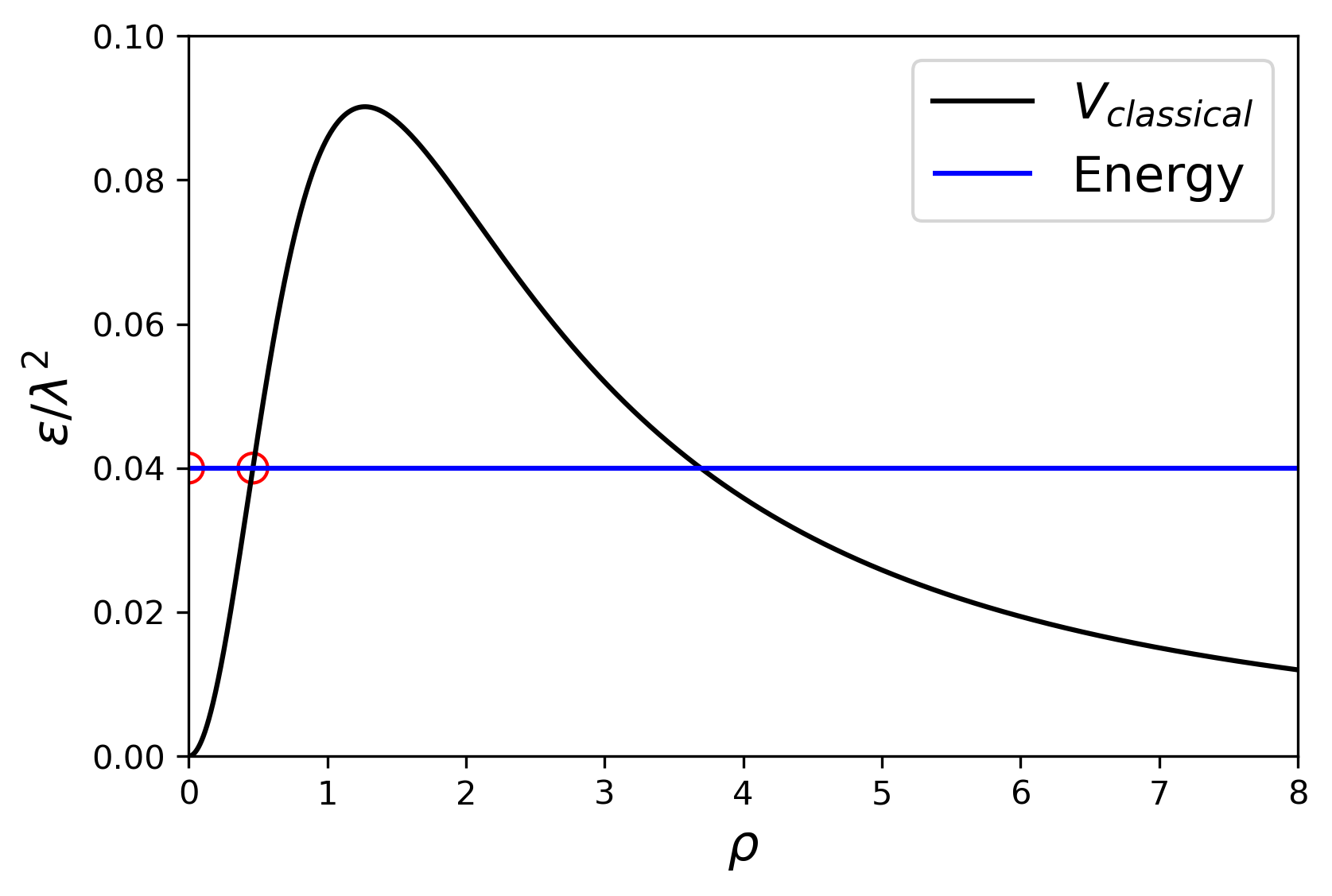}
\caption{}
\end{subfigure}
\begin{subfigure}{0.49\linewidth}
\includegraphics[width=\linewidth]{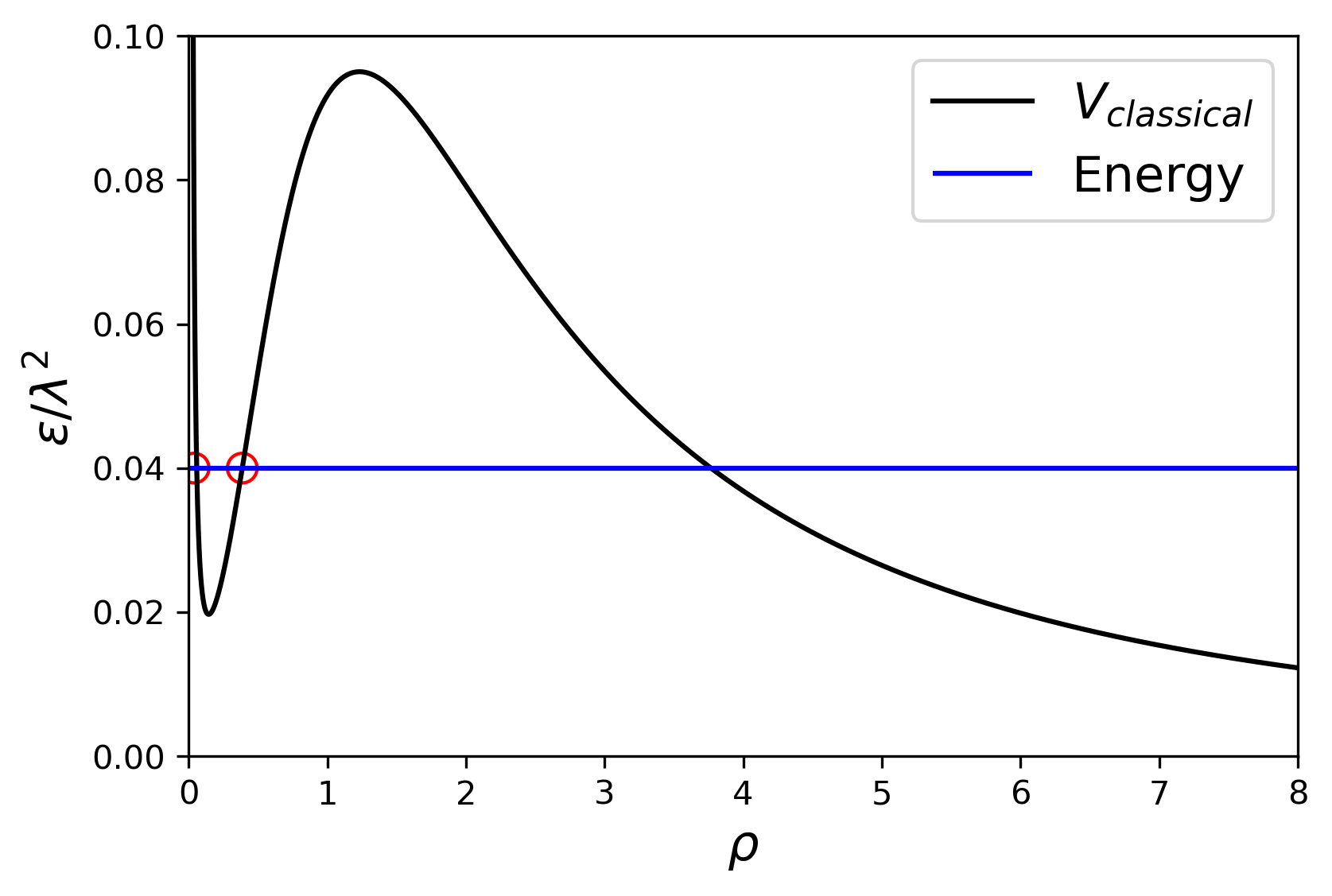}
\caption{}
\end{subfigure}
\caption{Examples of the effective radial potential $V_{\rm cl}(\rho)$ (black curves) for three different values of the electron's angular momentum:\ 
 (a) \(M/\lambda=-0.01\); (b) \(M/\lambda=0\); (c) \(M/\lambda=0.01\). 
The blue line shows an example radial energy of $400 E_0$ for the case $\lambda=100$.  The turning points for a particle of that radial energy that starts its radial motion within the potential well are shown as red circles.}
\label{Fig classic potentials for orbit}
\end{figure}

The limits on the radial motion for a particle of (dimensionless) energy $\epsilon$, i.e.\ the turning points, are given by the points where $V_{\rm cl} = \epsilon$.  Examples of these for three different values of $M/\lambda$ are shown in Fig.~\ref{Fig classic potentials for orbit}.
\begin{figure}
\centering
\begin{subfigure}{0.51\linewidth}
    \includegraphics[width=\linewidth]{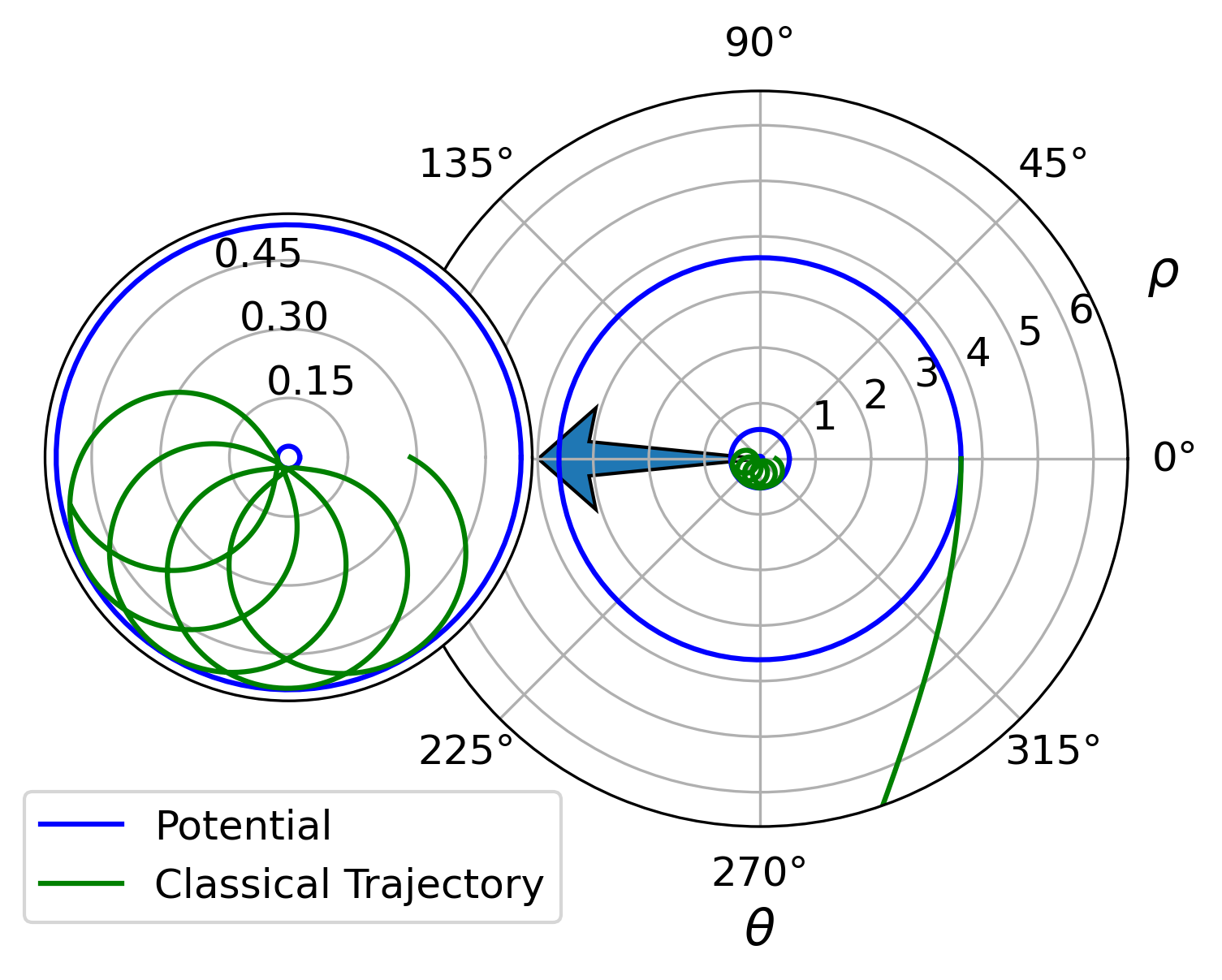}
    \caption{}
\end{subfigure}
\begin{subfigure}{0.49\linewidth}
    \includegraphics[width=\linewidth]{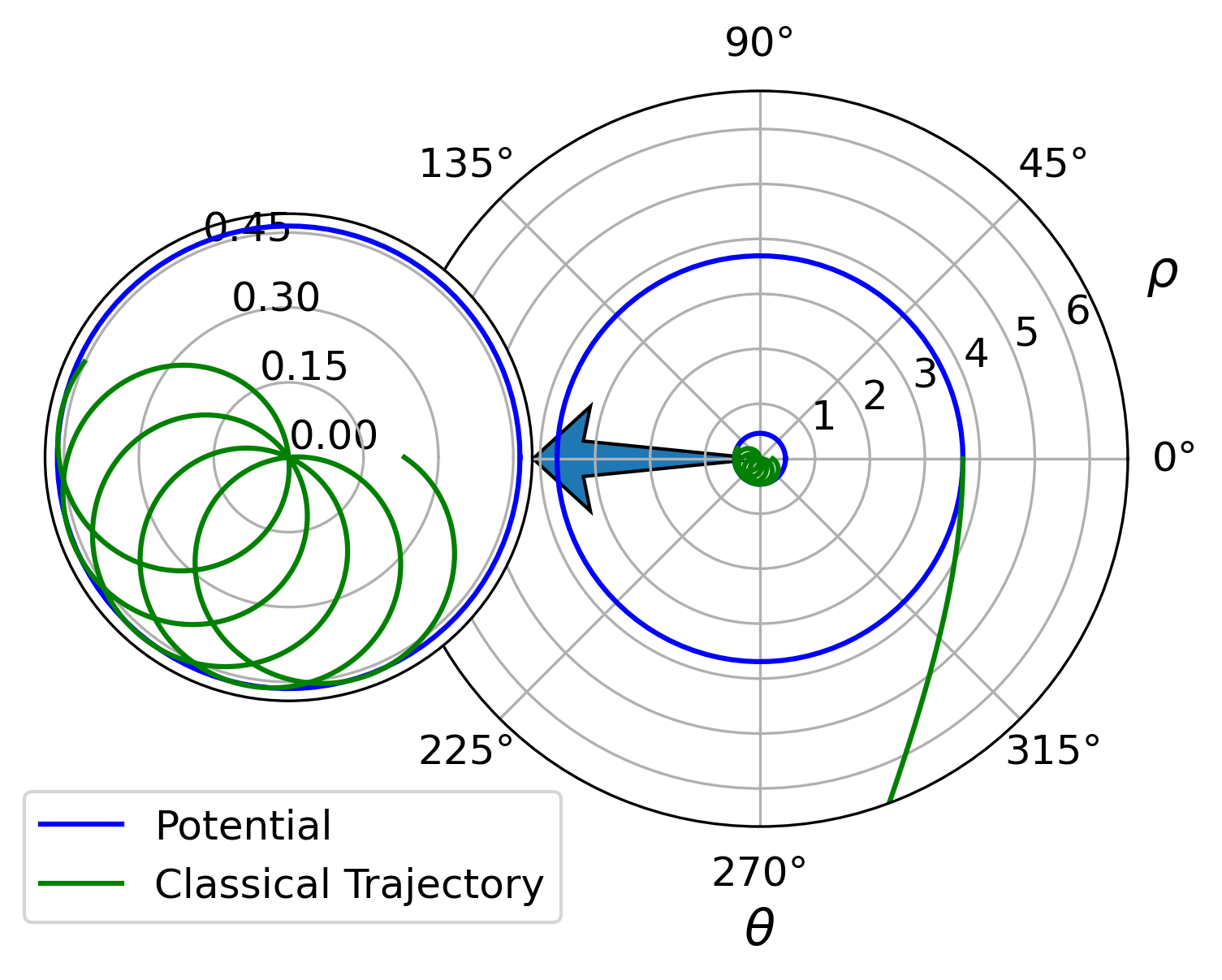}
    \caption{}
\end{subfigure}
\begin{subfigure}{0.49\linewidth}
    \includegraphics[width=\linewidth]{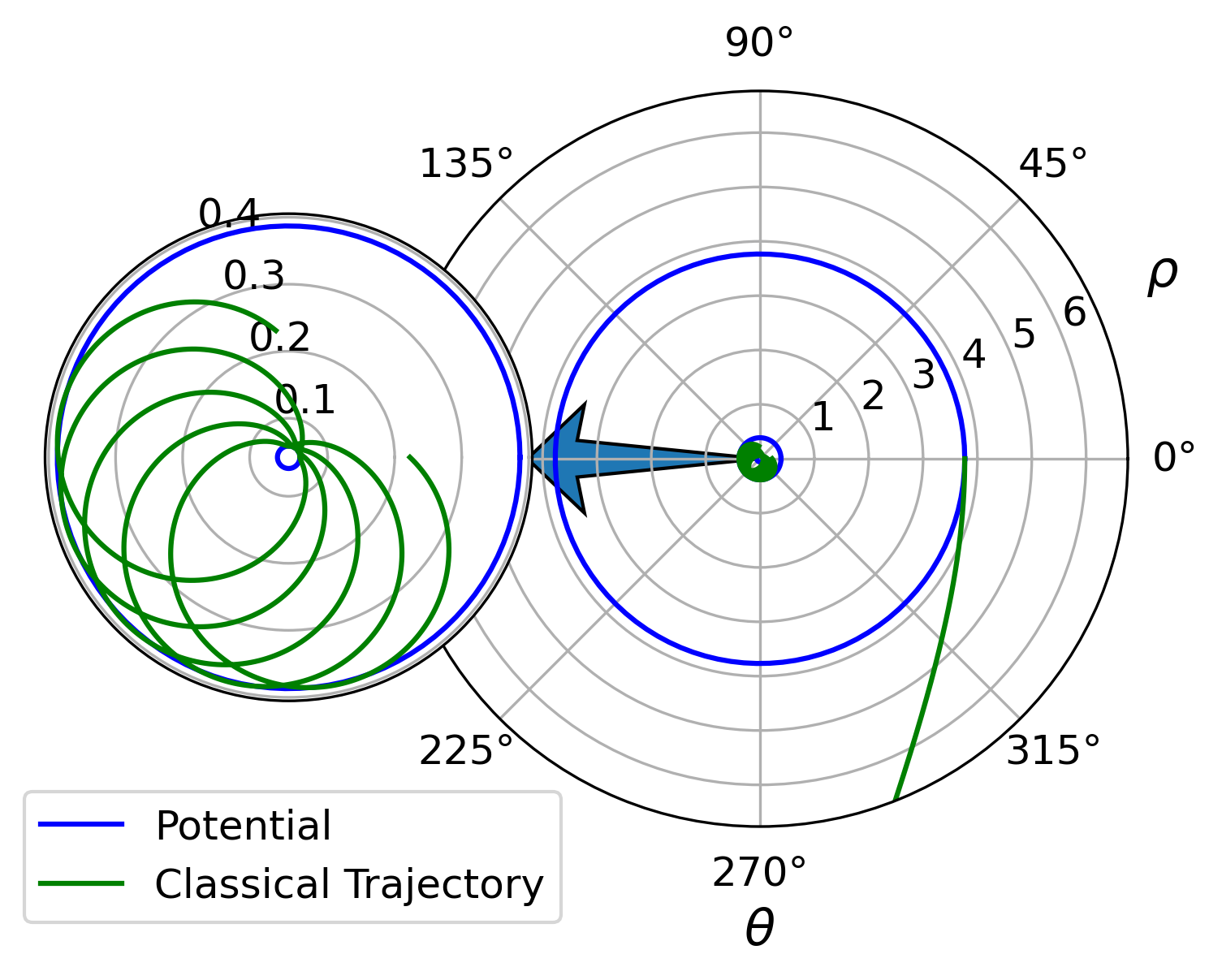}
    \caption{}
\end{subfigure}
\caption{Selected orbits corresponding to the potentials seen in Fig.~\ref{Fig classic potentials for orbit} (green curves), with one orbit starting within the well and one starting outside the well for the same value of the electron's energy:\ (a) \(M/\lambda=-0.01\); (b) \(M/\lambda=0\); (c) \(M/\lambda=0.01\). The blue lines labelled `Potential' indicate the radii at which \(V_\text{cl}(\rho)=\epsilon\), i.e.\ the radial turning points. The inset is a zoomed-in view of the region inside the potential well, where all bound states are located. The particle that starts outside the well moves away from the monopole and is unbound.}
\label{Fig classical all classical orbits}
\end{figure}

To describe the orbit in the 2D plane we must combine this radial information with the azimuthal motion given by (\ref{hameq2}),
\begin{equation}
\dot \phi = \frac{1}{m^{*} r^2} \left( p_\phi - q_e r A_\phi \right).
\end{equation}
Examples of the resulting classical orbits are shown in Fig.~\ref{Fig classical all classical orbits}.  The bound orbits are of two broad types:\ those in which individual orbit cycles enclose the circle corresponding to the inner turning point (positive $M/\lambda$ --- see panel (c)), and those in which they do not (negative $M/\lambda$ --- see panel (a)).  A detailed analysis of the motion shows that the circulation of the orbits is the same for both positive and negative $M$, depending only on the sign of $\lambda$.  Panel (b) shows the boundary case between these, where the inner turning point is at the origin.  There are, in addition, scattering-state orbits where the particle begins outside the outermost turning point and is unbound.

\begin{figure}%plot of circular orbits
    \centering
    \begin{subfigure}[c]{0.49\textwidth}
    
        \centering
        \includegraphics[height=52mm,width=\textwidth]{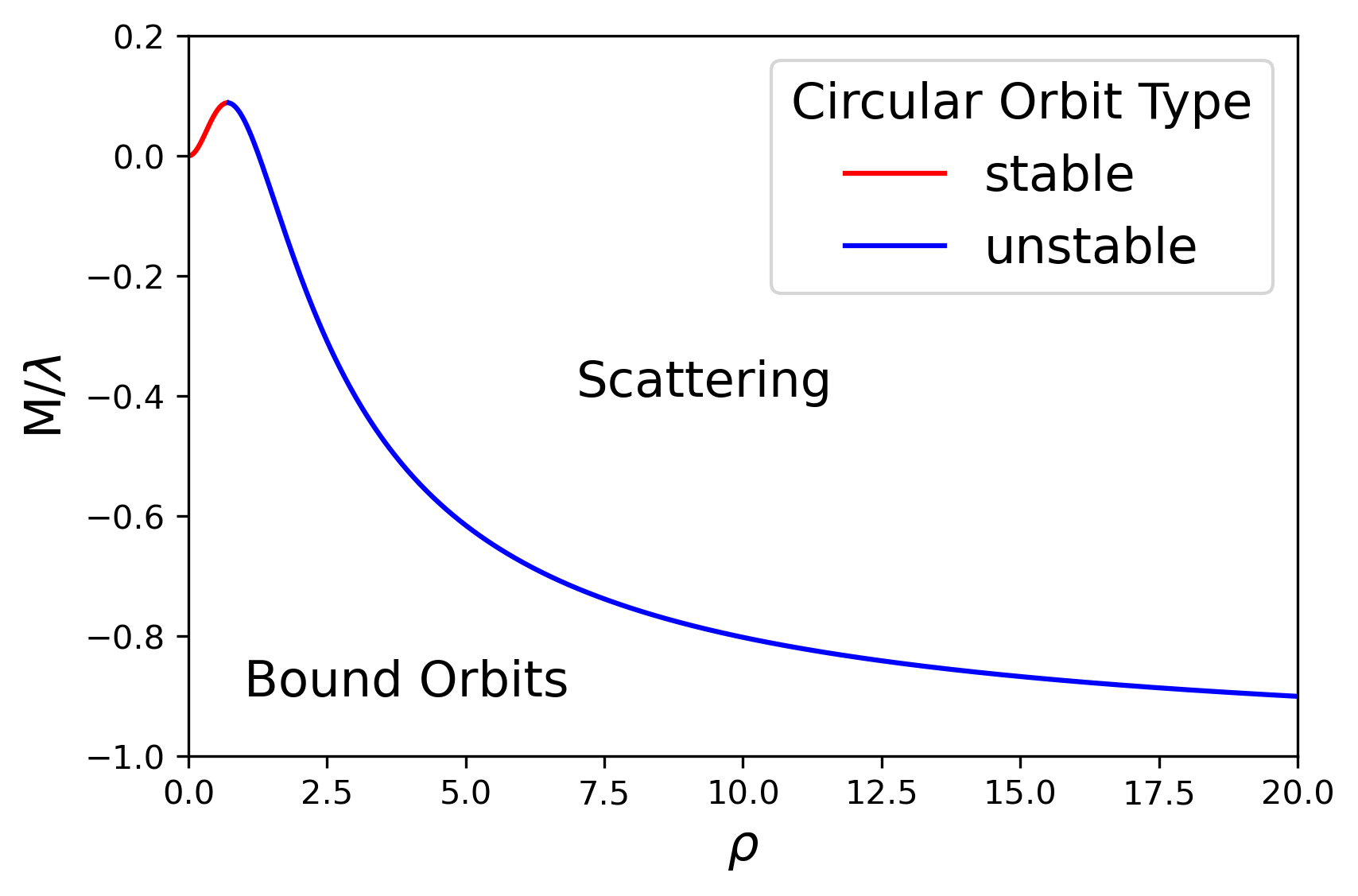}
        \caption{}
        \label{Fig classical circular orbits for a given value of M}
    \end{subfigure}
    \begin{subfigure}[c]{0.49\textwidth}
        \centering
        \includegraphics[height=50mm,width=\textwidth]{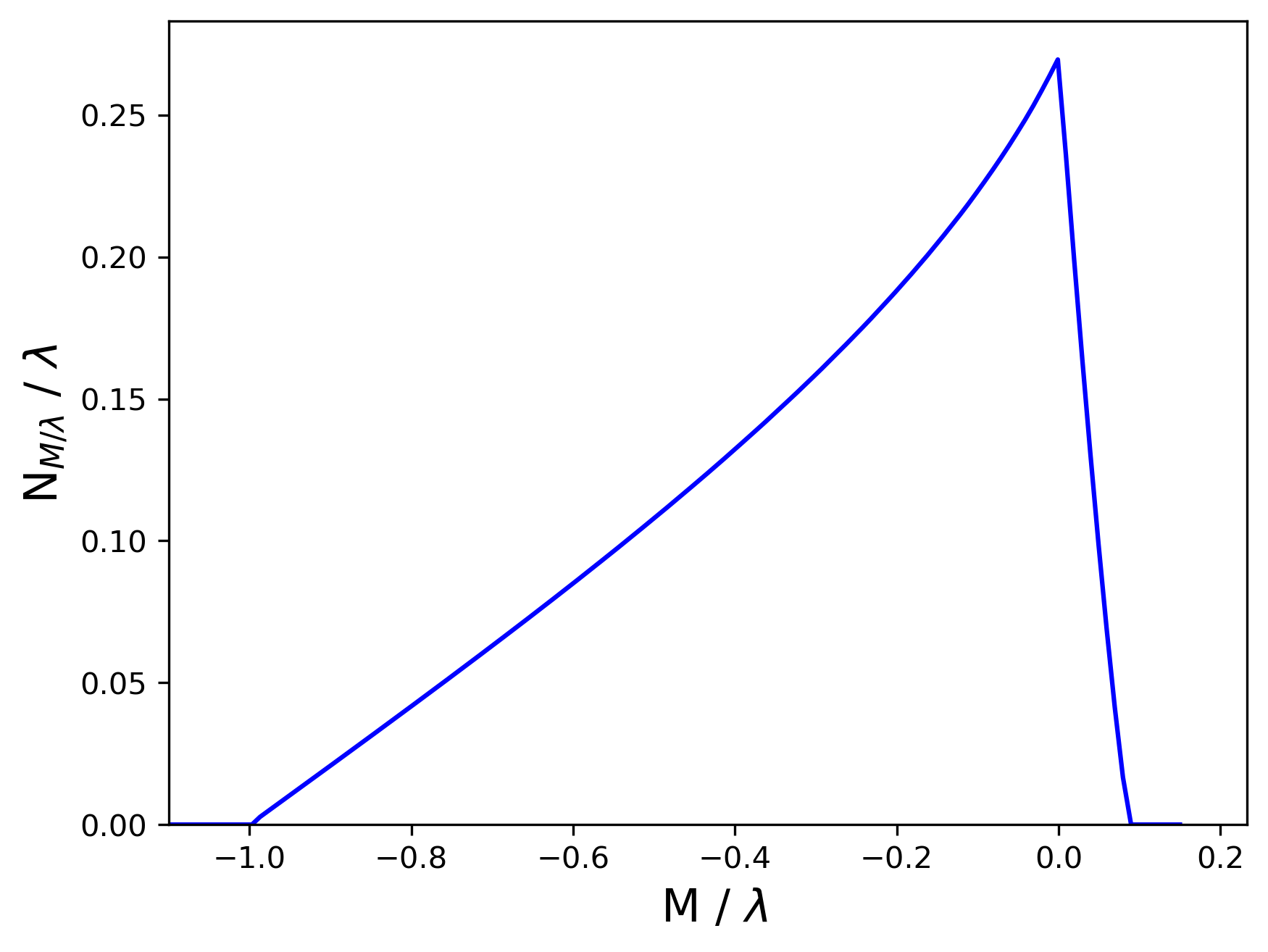}
        \caption{}
        \label{Fig classical Number of states for a given M}
    \end{subfigure}
\caption{(a) Relationship between angular momentum, $M$, and distance from the plane's origin, \(\rho\), for circular orbits. Above the curve all orbits are unbound; below the curve they are non-circular bound orbits (see Fig.~\ref{Fig classical all classical orbits}).  For values of \(M/\lambda>0\) there are two circular orbits; otherwise only one circular orbit is present. Some circular orbits are stable while others are unstable.  The stable circular orbits are found at the minimum of the effective potential for positive-valued \(M/\lambda\); the unstable ones are found at the maximum of the effective potential for all values of \(M/\lambda\). (b) The approximate number of bound states, scaled by \(\lambda\), as a function of \(M/\lambda\).  This curve is calculated by modelling the system as a simple harmonic oscillator --- see text for details.}
\end{figure}

There are also, of course, circular orbits --- i.e.\ orbits of time-independent radius.  These occur when the particle starts at a radius that corresponds to an extremum of the classical potential, and they can be classified by their stability under small perturbations. 
 Fig.~\ref{Fig classical circular orbits for a given value of M} shows the locations of those circular orbits in the $\rho$--$M/\lambda$ plane:\ the circular orbits separate the bound states from the scattering states.  (It might seem puzzling that there is only one branch of circular orbits for negative $M/\lambda$, whereas the radial effective potential always has an even number of turning points.  This is because the minima in the radial effective potentials for negative $M/\lambda$ --- see Fig.~\ref{Fig Classical Potentials} --- occur at zero energy, and thus correspond to a stationary electron rather than to a circular orbit.)

Even before solving the quantum version of this problem, which we shall do in later sections, we may make a rough estimate of whether the monopole will be strong enough to host bound states when the electron is described quantum mechanically.  Approximating the confining potential as that of a harmonic oscillator, we may estimate the number of quantum bound states for a chosen \(M/\lambda\) as
\begin{equation}
    \centering
    N_{M/\lambda} \approx w_{1/2} \lambda \sqrt{\frac{\epsilon_{1/2}}{2}},
    \label{Eq classical number of states}
\end{equation}
where $w_{1/2}$ is the width of the potential well at half of its depth and $\epsilon_{1/2}$ is the dimensionless energy there.
The derivation of this equation for \(N_{M/\lambda}\) can be found in Appendix~\ref{appendix derivation of N}; the function is plotted in Fig.~\ref{Fig classical Number of states for a given M}. The total number of bound states for any \(\lambda\),
\begin{equation}
    N_{b} = \sum_M N_{M/\lambda},
\end{equation}
is given by the area under the curve, which we numerically evaluate to be approximately 
\begin{equation}
    N_b\approx0.13\,\lambda^2.
    \label{classical_number_of_bound_states}
\end{equation}
This value is independent of $D$, the distance between the monopole and the plane, because the characteristic magnetic length scale is proportional to $D$ for a fixed value of the monopole charge \(Q_m\). 
It suggests a threshold monopole strength of $\lambda \approx 2.77$ for the occurrence of a single bound state; we shall compare this prediction with that of more accurate quantum treatments below.

\subsection{Semi-classical solution:\ Bohr-Sommerfeld quantisation}
 Bohr-Sommerfeld quantisation is applicable to periodic motion with one degree of freedom \cite{RefWorks:10,RefWorks:11,RefWorks:14}. This technique works well for circular orbits; on the other hand, should the particle experience turning points --- such as those indicated in Fig.~\ref{Fig classic potentials for orbit} --- a correction term has to be included to take these into account. The corrected Bohr-Sommerfeld quantisation condition is
 \begin{equation}
	\centering
	\oint p_s ds = 2\pi\left(n+\gamma\right)\hbar,
	\label{Eq Bohr Sommefield corrected}
\end{equation}
where \(s\) is the coordinate conjugate to the momentum, \(p_s\), being quantised, and $n$ is an integer. The correction term $\gamma$ can be evaluated by several different means \cite{RefWorks:16,RefWorks:5,RefWorks:20,RefWorks:24}. When \(\gamma=0\) the original form of the Bohr-Sommerfeld quantisation condition is recovered \cite{RefWorks:10,RefWorks:11,RefWorks:14,RefWorks:21,RefWorks:22,RefWorks:23,RefWorks:4}. Since the orbital and radial motions are separable, we can use the Bohr-Sommerfeld quantisation condition in a similar fashion as in Sommerfeld's study of the hydrogen atom \cite{RefWorks:4}.

For the angular momentum there is no correction term \cite{RefWorks:16,RefWorks:5,RefWorks:20,RefWorks:24}, and the Bohr-Sommerfeld quantisation condition is simply
\begin{equation}
	\centering
		\int_{0}^{2 \pi} p_\phi d\phi = 2 \pi M \hbar \hspace{6mm} \Longrightarrow  \hspace{6mm} p_\phi=M\hbar.
	\label{Eq bohr angular momentum}
\end{equation}
Here \(p_\phi\) is the orbital angular momentum and \(M\) is the corresponding angular momentum quantum number, which may be any integer (positive, negative, or zero).  The radial quantisation condition, on the other hand, requires a correction term; it reads:
\begin{equation}
	\centering
		\oint p_r dr = 2 \pi \left(n + \gamma\right) \hbar \hspace{6mm} \Longrightarrow \hspace{6mm} \int_{r_1}^{r_2} p_r dr =  \pi \left(n+\frac{1}{2}\right)\hbar,
	\label{Eq Bohr quantisation of rdial momentum.}
\end{equation}
where $r_1$ and $r_2$ are the radii of the two turning points, and the correction term \(\gamma=1/2\)  \cite{RefWorks:16,RefWorks:5,RefWorks:20,RefWorks:24}.
The expression for the total energy of the system is the same as the classical case, with the same canonical momentum, and thus the effective potential that is derived through Bohr-Sommerfeld quantisation is the same as in the classical case.
The dimensionless Bohr-Sommerfeld quantisation condition for the radial motion can thus be expressed as
\begin{equation}
	\centering
	\frac{1}{\pi}\int_{\rho_1}^{\rho_2}  \sqrt{\epsilon - V_\text{cl}(\rho)} \,d\rho = \left(n + \frac{1}{2}\right).
	\label{Eq Bohr quantisation for monopole}
\end{equation}
Here the quantum numbers are $n=0,1,2,\ldots$ and \(\rho_1\) and \(\rho_2\) are the dimensionless radii of the classical turning points, given by \(V_\text{cl}(\rho)=\epsilon\) for \(\rho\geq0\).

We shall return to the results of the Bohr-Sommerfeld quantisation after looking at the quantum mechanical solutions --- see Subsection~\ref{section energy spectrum and eigen functions}\ref{subsection energy spectrum}.

\section{Energy spectrum and eigenfunctions}\label{section energy spectrum and eigen functions}

\subsection{The quantum mechanical problem}
Promoting the momentum in (\ref{Eq hamiltonian}) to an operator, we obtain the Hamiltonian operator of the quantum problem in the position basis.  A definite-energy state of the particle is now described by a time-independent wave function \(\Psi(r, \phi)\) satisfying the time-independent Schr\"odinger equation
\begin{equation}
    \centering
    \frac{1}{2m^{*}}\left(\hat{{\bf p}}-q_e\mathbf{A}\right)^2 \Psi\left(r,\phi\right)=E\,\Psi\left(r,\phi\right),
    \label{tise2d}
\end{equation}
where \(\mathbf{A}=A_\phi(r)\hat{\mbox{\boldmath $\phi$}}\), with \(A_\phi(r)\) given by (\ref{Eq A dimesionfull}).
We seek separable solutions of the form \(\Psi\left(r,\phi\right)=G(r)\,\text{e}^{iM\phi}\), where \(G(r)\) is a function of $r$ and $M$ (an integer) is the angular momentum quantum number.  The equation for \(G(r)\) is derived in Appendix~\ref{appendix radial equation}; it can be written in terms of the dimensionless radius and energy scales given in (\ref{r_scale}) and (\ref{E_scale}) as follows:
\begin{equation}
    -\frac{d^2 G(\rho)}{d \rho^2} - \frac{1}{\rho} \frac{d G(\rho)}{d \rho} + V(\rho)G(\rho)=\epsilon \,G(\rho),
    \label{Eq dimensionless radial schrodinger equation}
\end{equation}
where \(V(\rho)\) is defined as:
\begin{equation} 
    V(\rho)=\frac{\lambda^2}{\rho^2}\left[\frac{M}{\lambda} +  \left(1 - \frac{1}{\sqrt{1 + \rho^2}}\right)\right]^2.
\end{equation} 
This coincides with the classical potential, \(V_\mathrm{cl}(\rho)\) in (\ref{eq classical scaled potential}); however, \(V(\rho)\) does not play the role of an effective potential for the radial part of the quantum problem, as we shall now see.

Using the substitution \(G(\rho)=\rho^{-1/2}\psi(\rho)\) we may derive a one-dimensional Schr\"odinger equation for \(\psi(\rho)\):
\begin{equation}
    \centering
    -\frac{d^2 \psi(\rho)}{d \rho^2} + V_\text{q}(\rho)\psi(\rho)=\epsilon\psi(\rho),
    \label{Eq Quantum one dimensional schroinger equation}
\end{equation}
with the potential \(V_\text{q}(\rho)\):
\begin{equation}
    \centering
    V_\text{q}(\rho)=\frac{\lambda^2}{\rho^2}\left(\frac{M}{\lambda} + \left[1 - \frac{1}{\sqrt{1 + \rho^2}}\right]\right)^2 - \frac{1}{4 \rho^2}.
\end{equation}

There are, as far as we know, no exact solutions to (\ref{Eq Quantum one dimensional schroinger equation}) for this form of the potential, so we used approximate methods to evaluate $\psi(\rho)$ both analytically and numerically. Our analytical treatment uses the WKB approximation, and is presented in subsection~\ref*{section energy spectrum and eigen functions}\ref{sub section wkb approximation}; our numerical treatment uses the finite difference method, and is presented in subsection~\ref*{section energy spectrum and eigen functions}\ref{subsection numerical solution}. 

\begin{figure}
    \centering
    \begin{subfigure}{0.49\linewidth}
    \centering
    \includegraphics[width=\linewidth]{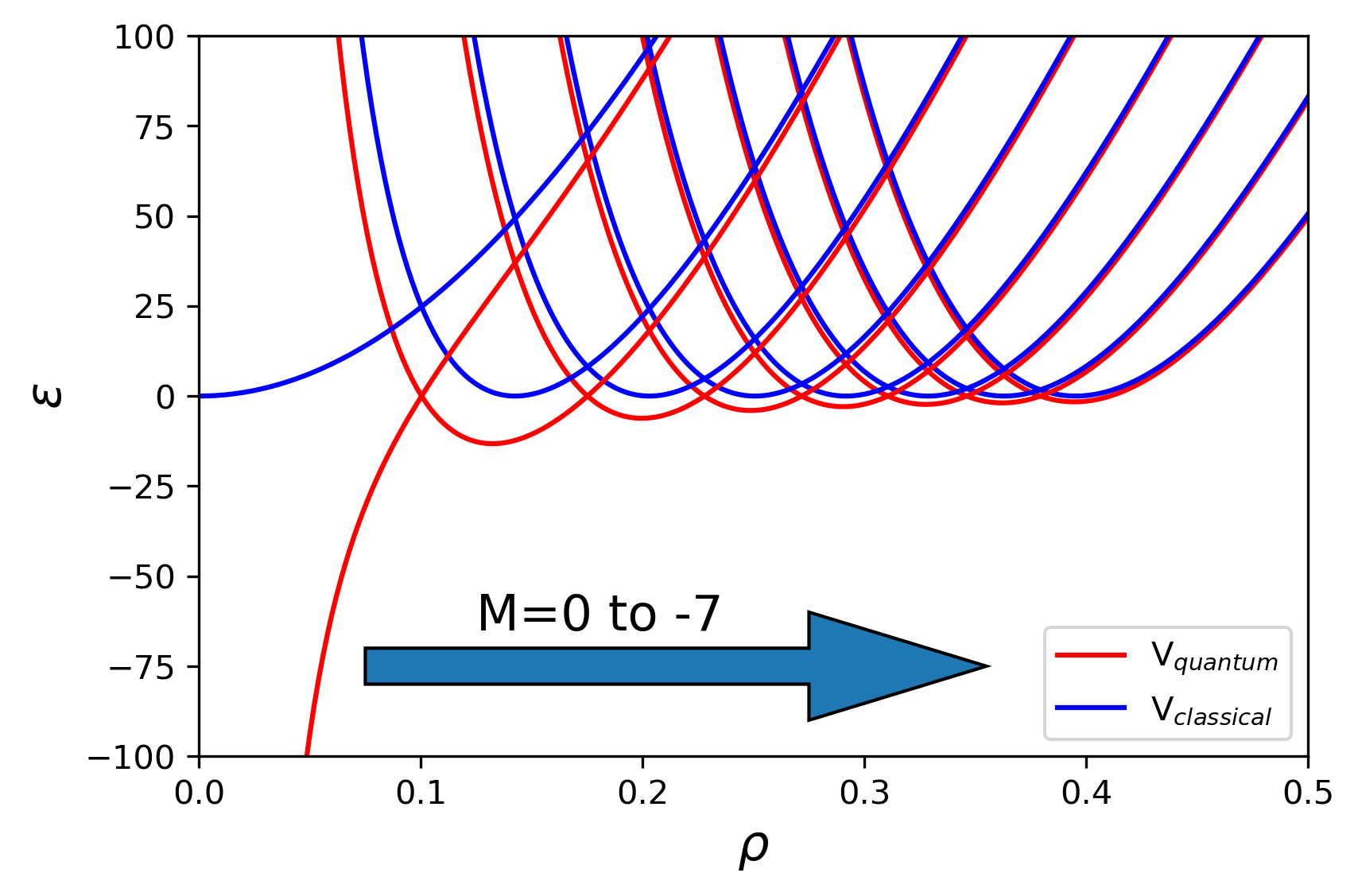}
    \caption{}
    \end{subfigure}
    \begin{subfigure}{0.49\linewidth}
    \centering
    \includegraphics[width=\linewidth]{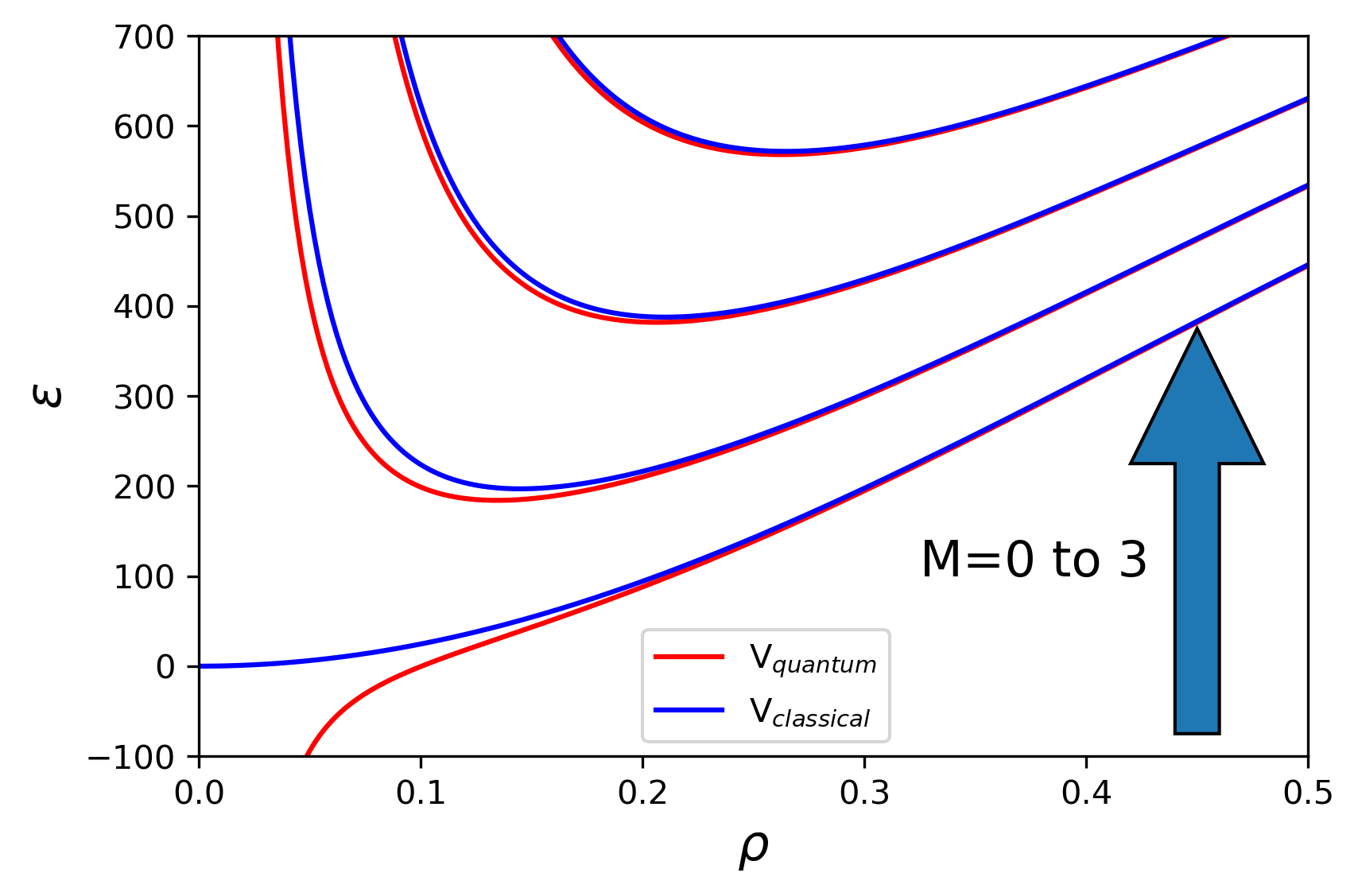}
    \caption{}
    \end{subfigure}

    \caption{Comparison of the classical and quantum potentials for (a) $-7 \leqslant M \leqslant 0$; (b) $0 \leqslant M \leqslant 3$. In both cases, we see the potentials converging as \(\rho\) is increased. The greatest discrepancy between the two potentials is for \(M=0\):\ as \(\rho \to 0\) the quantum potential tends to \(-\infty\) while the classical potential goes to \(0\). The reason for this is the additional term \(-1/(4\rho^2)\) in the quantum potential compared to the classical one.}
    \label{Fig comparison of the quantum and classical potentials.}
    \end{figure}

The difference between the potentials \(V_\text{cl}(\rho)\) and \(V_\text{q}(\rho)\) is the additional term \(-1/(4\rho^2)\) found in \(V_\text{q}(\rho)\); the effect of this term may be seen in Fig.~\ref{Fig comparison of the quantum and classical potentials.}. The larger the absolute value of \(M\), the closer \(V_{\rm cl}\) will be to \(V_{\rm q}\), and this convergence is faster for positive values of \(M\) than for negative ones. For \(M=0\) the quantum potential diverges to minus infinity as \(\rho \to 0\), while the classical one goes to a finite constant.

To understand the origin of this discrepancy, let us consider two ways in which the effective one-dimensional Schr{\"o}dinger equation might be derived:

The first is to start with the two-dimensional classical Hamiltonian in Cartesian co-ordinates, transform it into polar coordinates using \(x=r\cos\phi\) and \(y=r\sin\phi\),
\begin{equation}
    \centering
    H_\text{polar}=\frac{1}{2m}\left[P_r^2 + \frac{P_\phi^2}{r^2}\right] +V(r,\phi),
    \label{Eq eigenvalues classical hamiltonian polar coordinates}
\end{equation}
and then promote the momenta in (\ref{Eq eigenvalues classical hamiltonian polar coordinates}) to operators to form a position-basis Schr\"odinger equation for a wave function \(\psi_\text{c}\).  Using this method, we obtain:
\begin{equation}
    \centering
    -\frac{\hbar^2}{2 m}\left[\frac{\partial^2}{\partial r^2}+\frac{1}{r^2}\frac{\partial^2}{\partial \phi^2}\right]\psi_\text{c}(r,\phi) +V(r,\phi)\psi_\text{c}(r,\phi)=E\,\psi_\text{c}(r,\phi).
    \label{Eq eigenvalues schrodinger equaion in polar coordinates derived fom the classicla hamiltonian}
\end{equation}

The second method is to interchange these two steps, first writing a two-dimensional Schr{\"o}dinger equation in Cartesian co-ordinates and then changing to polar co-ordinates by transforming the variables in the Laplacian. This results in a different form of the Schr\"odinger equation, now written for a wave function \(\psi_\text{q}\):
\begin{equation}
    \centering
    -\frac{\hbar^2}{2m}\left[\frac{\partial^2}{\partial r^2} + \frac{1}{r}\frac{\partial}{\partial r}+\frac{1}{r^2}\frac{\partial^2}{\partial \phi^2}\right]\psi_\text{q}(r,\phi) + V(r,\phi)\psi_\text{q}(r,\phi)=E\,\psi_\text{q}(r,\phi).
    \label{Eq eigenvalues schrodinger equation in polar coordinates derived from the schrodinger equation}
\end{equation}

The difference between equation (\ref{Eq eigenvalues schrodinger equaion in polar coordinates derived fom the classicla hamiltonian}) and (\ref{Eq eigenvalues schrodinger equation in polar coordinates derived from the schrodinger equation}) is the additional first-radial-derivative term in (\ref{Eq eigenvalues schrodinger equation in polar coordinates derived from the schrodinger equation}).  The correct method is the second one \cite{Paz2000,Paz2000(1)}, i.e.\ the one in which this additional term is obtained.

\subsection{WKB approximation}\label{sub section wkb approximation}
The WKB approximation obtains an analytic solution of the Schr\"odinger equation as an asymptotic series expansion, the terms of which are derived from the energy-momentum relation of a classical particle moving in the quantum potential $V_q(\rho)$. Here we go to the second order in this expansion; higher orders can be chosen for more accuracy, but at an increased computational cost. Using the methods described in refs.\ \cite{RefWorks:1,RefWorks:2,RefWorks:3}, we obtain an implicit equation for the  energy eigenvalues in the WKB approximation:
\begin{equation}
	\centering
\int_{\rho_1}^{\rho_2}p_n(\rho)d\rho=\left(n+\frac{1}{2}\right)\pi,
	\label{Eq WKB Energy eigenvalues}
\end{equation}
where \(p_n(\rho)=\sqrt{ \epsilon_n-V_\text{q}(\rho)}\), $\epsilon_n$ being the energy eigenvalue of the eigenstate with quantum number $n$.
The WKB approximation produces a piecewise wave function \(\psi^{\text{WKB}}_n(\rho)\):
\begin{equation}
	\centering
	\psi^{\text{WKB}}_n(\rho)=
	\left\{ \begin{array}{lll}
		\displaystyle \frac{C_n}{\sqrt{\left|p_n(\rho)\right|}}\exp\left[\int_{\rho}^{\rho_1}p_n(\rho')d\rho'\right]& & \rho<\rho_1, \\
		& & \\
		\displaystyle \frac{C_n}{\sqrt{\left|p_n(\rho)\right|}}\cos\left[\int_{\rho_1}^{\rho}p_n(\rho')d\rho'-\frac{\pi}{4}\right] & & \rho_1<\rho<\rho_2, \\
		& & \\
		\displaystyle \frac{(-1)^nC_n}{\sqrt{\left|p_n(\rho)\right|}}\exp\left[-\int_{\rho_2}^{\rho}p_n(\rho')d\rho'\right] & & \rho_2<\rho<\rho_3,\\
		& & \\
		\displaystyle \frac{2(-1)^nC_n}{\sqrt{\left|p_n(\rho)\right|}}\exp(J_n)\cos\left[-\int_{\rho_3}^{\rho}p_n(\rho')d\rho'-\frac{\pi}{4}\right] & \qquad \qquad & \rho_3<\rho,
	\end{array} \right.
	\label{Eq WKB Psi piecwise function}
\end{equation}
where
\begin{equation}
	\centering
		J_n= -\int_{\rho_2}^{\rho_3}p_n(\rho)d\rho,
	\label{Eq WKB tunneling constant}
\end{equation}
and \(\rho_1\), \(\rho_2\), and \(\rho_3\) are the classical turning points at which \(\epsilon_n=V_\text{q}(\rho_i)\) for \(i=1,2,3\). At these turning points \(p_n(\rho)=0\), which causes a singularity in the wave function there.
The normalisation coefficient, \(C_n\), is given by
\begin{equation}
    \centering
    C_n = \left(\int_{\rho_1}^{\rho_2}\frac{1}{p_n(\rho)}\cos^2\left[\int_{\rho_1}^{\rho}p_n(\rho') d\rho' -\frac{\pi}{4}\right]d\rho\right)^{-\frac{1}{2}};
\end{equation}
here we have approximated the normalisation volume as the width of the well at \(\epsilon_n\), since the majority of the bound wave function's amplitude is within the well.
We present the results obtained using this WKB approximation in subsection \ref*{section energy spectrum and eigen functions}\ref{subsection energy spectrum}; first, though, we describe the numerical finite difference method used to obtain our numerical solutions.

\subsection{Numerical solution}\label{subsection numerical solution}
To obtain numerical solutions of the Schr{\"o}dinger equation for this problem we use the finite difference method, an approach in which the equation is diagonalised in a discrete basis to solve for \(\psi(\rho)\). The matrix equation that we use for this is
\begin{equation}
    \centering
    \left[\frac{1}{h^2}\boldsymbol{T}+\boldsymbol{U}\right]\psi(\rho) = \epsilon\,\psi(\rho),
    \label{Eq Finite diagonalised schrodinger equation}
\end{equation}
where \(\boldsymbol{T}\) is the tridiagonal matrix corresponding to the discretisation of the radial derivatives, \(\boldsymbol{U}\) is the diagonal matrix obtained by discretising the quantum potential \(V_\text{q}(\rho)\), and \(\psi(\rho)\) is a vector of length \(N\). The step size, \(h\), is given by \(h=\frac{b-a}{N-1}\), where $a \leqslant \rho \leqslant b$ is the interval on the radial line for which a solution is sought.

The accuracy of this numerical technique is controlled by the step size $h$:\ a small step size increases the accuracy, but also increases the computational power required to evaluate the solutions. A limiting factor for this particular method is the finite range of \(\rho\), which effectively means that the particle is enclosed in an infinite square well with a width $b-a$. On the one hand, a small step-size \(h\) introduces a cut-off in the spectrum of eigenvalues \(\epsilon\), and we can trust only eigenvalues that are small compared to this cut-off. On the other hand, the finite radial range \(L = b-a\) means that we obtain a discrete instead of a continuous spectrum of eigenvalues; thus, when discussing the energy-dependence of some property, our resolution will be limited accordingly. These issues can be addressed by decreasing the step-size \(h\) and increasing the width \(L\); this, however, increases the dimension of the matrices, \(N = L/h\), which must be kept within the limits of our computers. We have used \(N=4000\), \(a=0\), and \(b=20\), allowing us to compute wave functions and eigenvalues with good enough accuracy to make comparisons with WKB results. In particular, selecting the numerical eigenfunctions that have significant weight within the potential well reveals the quasi-bound states.

The results of the finite difference method can be found in subsection~\ref*{section energy spectrum and eigen functions}\ref{subsection energy spectrum}, where we compare the results of our different analytical and numerical approaches to the quantum-mechanical problem.

\subsection{Energy spectrum}\label{subsection energy spectrum}

\begin{figure}
\begin{center}\includegraphics[width=0.9\textwidth]{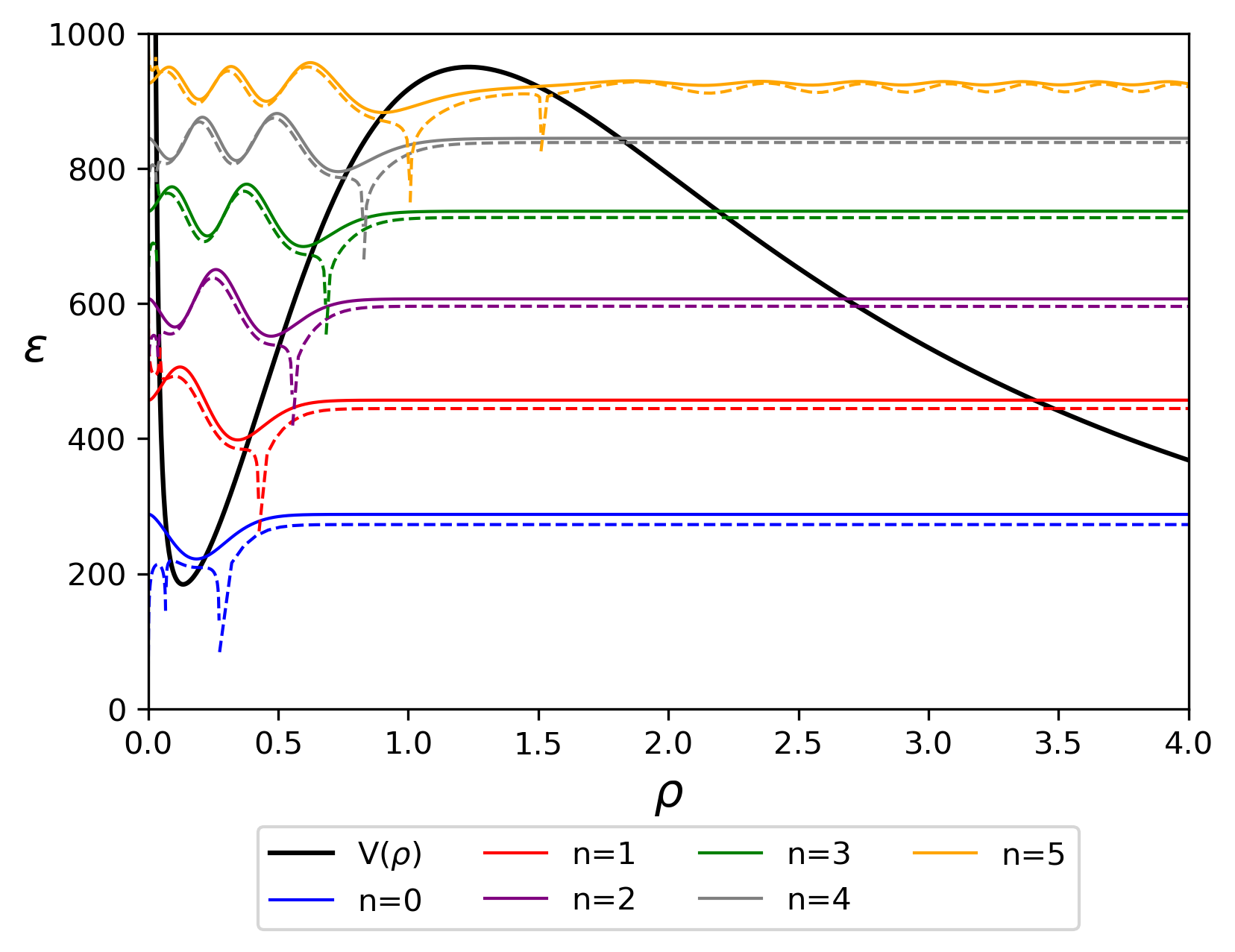}\end{center}
\caption{Quasi-bound state wave functions calculated numerically with the finite difference method (solid coloured curves) and  derived using the WKB approximation (dashed curves), offset vertically by their respective energy eigenvalues.  The dashed curves' singularities at the classical turning points, where \(\epsilon=V_{\rm q}(\rho)\), are artifacts of the WKB approximation. The quasi-bound solutions are for the effective potential \(V_{\rm q}(\rho)\) with $M=1$ and \(\lambda=100\) (solid black curve).}
    \label{Fig Finite WKB and finite wavefunction plots}
\end{figure}

\begin{figure}
    \centering
    \begin{subfigure}{0.49\textwidth}
    \includegraphics[width=\textwidth]{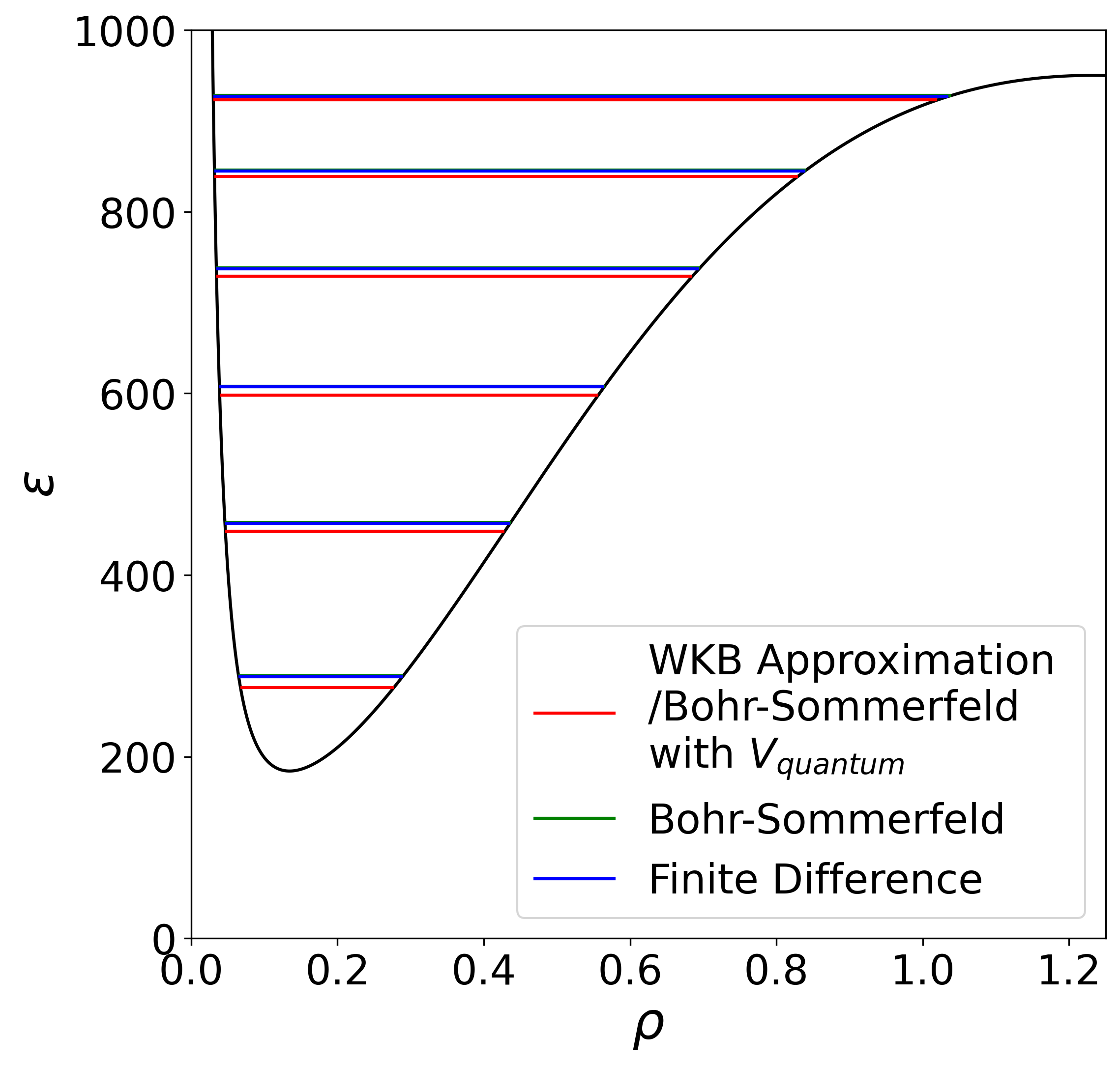}
    \caption{}
    \label{Fig Eigenvalues eigenvalues WKB bohr and fintie}
    \end{subfigure}
    \begin{subfigure}{0.49\textwidth}
    \includegraphics[width=\textwidth]{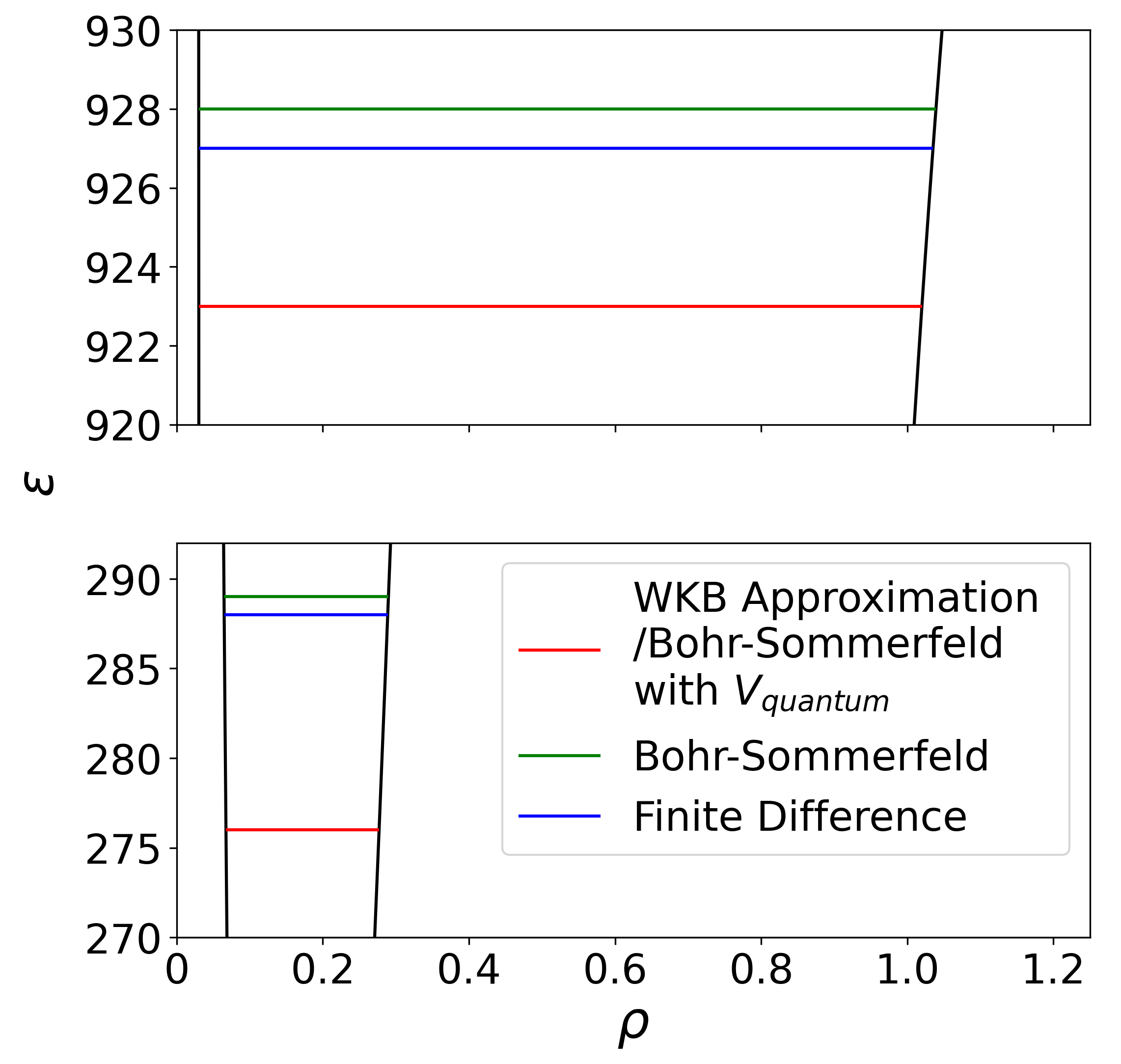}
    \caption{}
    \label{Fig Eigenvalues bohr with quantum potential}
    \end{subfigure}
    \caption{Comparison of eigenvalues for the WKB approximation, Bohr-Sommerfeld quantisation, and finite difference method. The Bohr-Sommerfeld energies (green) are calculated using \(V_{\rm cl}\). The Bohr-Sommerfeld quantisation evaluated using \(V_{\rm q}\) produces the same eigenvalues as the WKB approximation (red). (a) The full spectrum of the quasi-bound states for \(M=1\) and \(\lambda=100\). (b) Magnified view of the \(n=0\) and \(n=5\) states from (a).}
    \label{Fig energy eigenvalues}
\end{figure}

\begin{figure}
\begin{subfigure}{0.49\linewidth}
    \centering
    \includegraphics[width=\textwidth,height=60mm]{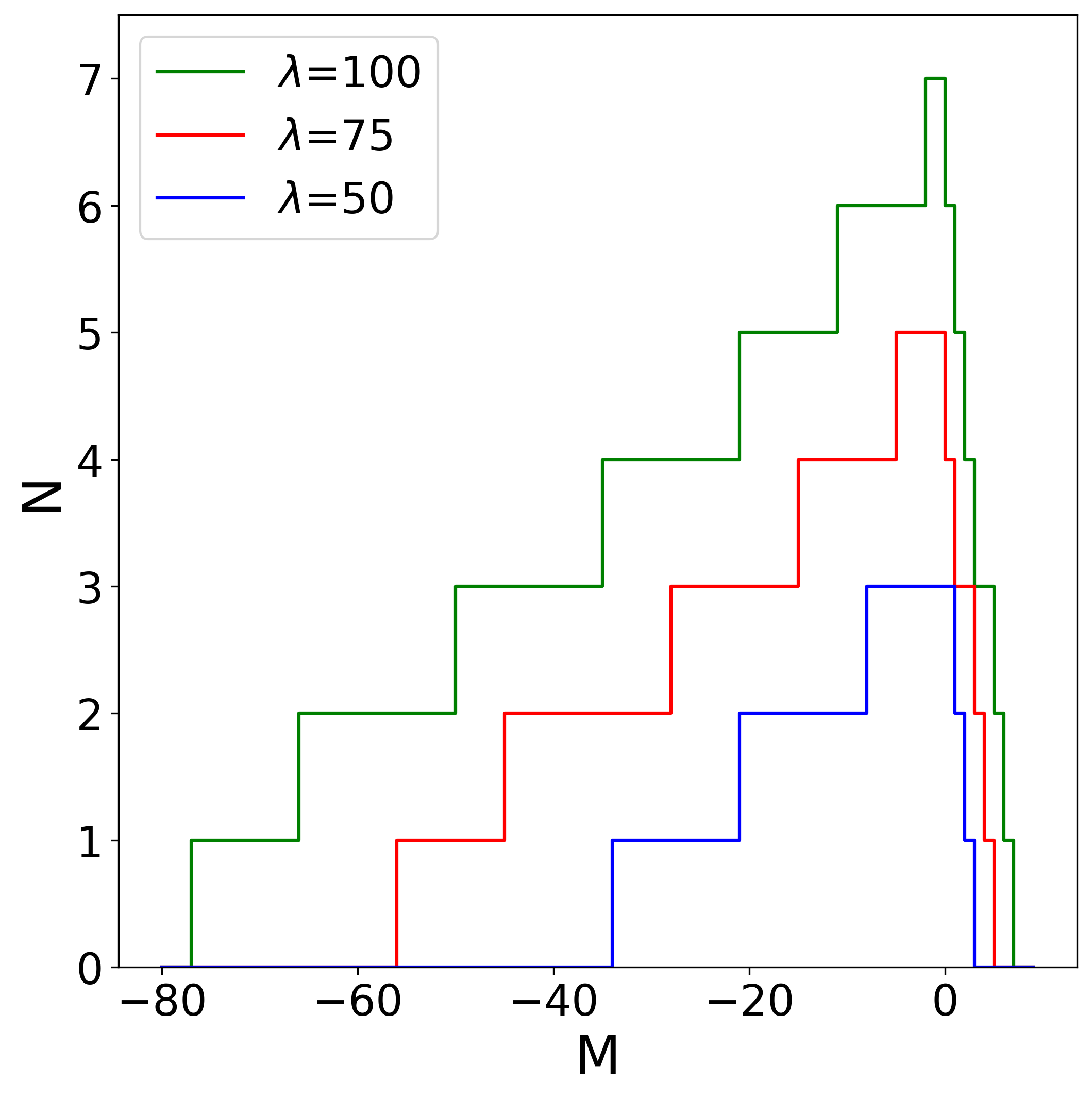}
    \caption{}
    \label{Fig Eigenvalues number of bound states}
\end{subfigure}
\begin{subfigure}{0.49\linewidth}
    \centering
    \includegraphics[width=\textwidth,height=60mm]{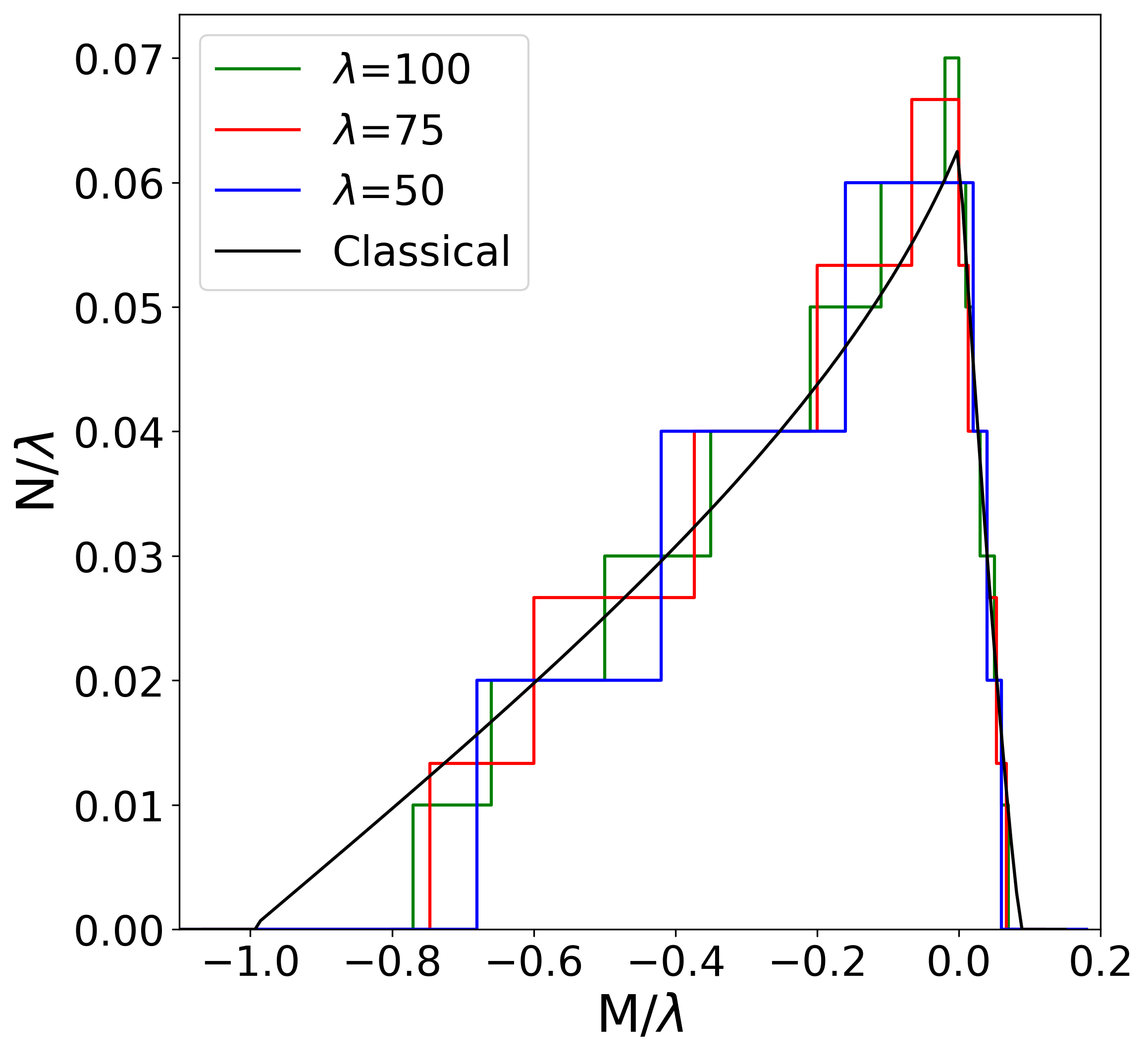}
    \caption{}
    \label{Fig comparison of quantum and classical}
\end{subfigure}
\caption{(a) Number of quasi-bound states as a function of $M$ for three different values of \(\lambda\). These curves are evaluated numerically using the finite difference method; the WKB approximation gives very similar results. (b) Number of quasi-bound states as a function of $M$ with both axes scaled by \(\lambda\) (coloured curves).  The estimate based on the harmonic approximation --- see Fig.~\ref{Fig classical Number of states for a given M} --- is shown, scaled by the phenomenological factor \(4.3\), for comparison (black curve).}
\end{figure}
The quasi-bound state wave functions for the case $M=1$, $\lambda=100$ are shown in Fig.~\ref{Fig Finite WKB and finite wavefunction plots}:\ the solid black curve is the effective potential $V_{\rm q}(\rho)$, the solid coloured curves are the wave functions evaluated numerically using the finite difference method, and the dashed coloured curves are the results of the WKB approximation.  (We call the states quasi-bound because, due to quantum tunnelling, the wave functions are not actually zero in the region outside the potential barrier.)  The results of our analytical methods, the WKB approximation and Bohr-Sommerfeld quantisation, become more similar to the numerical results as the quantum number $n$ increases.  If we use the potential \(V_\text{\rm q}(\rho)\) in the Bohr-Sommerfeld quantisation condition, we find that the energy eigenvalues are the same as the WKB results to five decimal places, as seen in Fig.~\ref{Fig energy eigenvalues}.

In the classical model, a bound state (which in that case is strictly bound) can be found for arbitrarily large negative $M$.  In the quantum model, by contrast, we find that for large negative values of \(M\), where the well is shallow and wide, no quasi-bound states are present.
Fig.~\ref{Fig Eigenvalues number of bound states} shows the numerically obtained number of bound states as a function of $M$ for three different values of $\lambda$; Fig.~\ref{Fig comparison of quantum and classical} shows these results scaled by $\lambda$ and compared with the semiclassical estimate based on a harmonic-oscillator approximation to the classical effective potential.  In all cases the largest number of bound states occurs at \(M=0\).

The total number of bound states that can be found for the quantum solution, \(N_{\rm bq}\), is given by the area under each plot in Fig.~\ref{Fig Eigenvalues number of bound states}.  We compute this to be approximately
\begin{equation}
    N_{\rm bq}=\sum_M N_M \approx 0.03\,\lambda^2.
    \label{Eq sum N/m}
\end{equation}

\subsection{Minimum Monopole Charge Required for a Single Quasi-Bound State}
\begin{figure}
    \centering
        \includegraphics[width=0.9\textwidth]{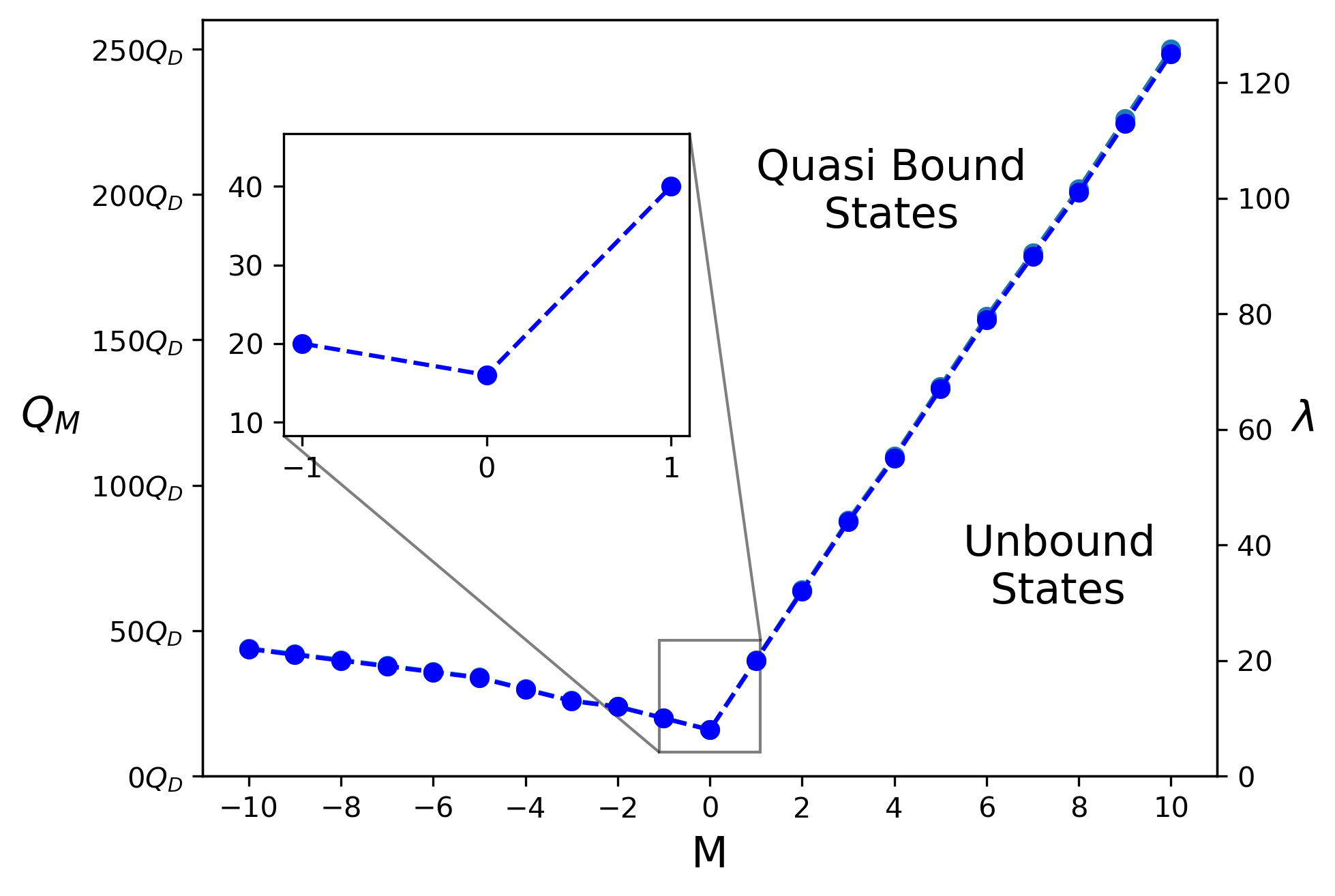}
        \caption{The minimum values of the dimensionless monopole strength $\lambda$ and the associated dimensionful strength \(Q_m\) required to produce a single quasi-bound state,  as functions of the angular momentum quantum number $M$. These values have been evaluated numerically using the finite difference method; the WKB approximation gives very similar results.}
        \label{Fig min lamba and qm}
\end{figure}

For a given angular momentum $M$ there is a threshold value of the dimensionless monopole strength \(\lambda\) for the appearance of the first quasi-bound state; this threshold is plotted in Fig.~\ref{Fig min lamba and qm}.  The weakest monopole that can produce a quasi-bound state is one of strength \(Q_m \sim 18Q_D\), and the quasi-bound state first occurs at an angular momentum $M=0$.

The physical meaning of the quasi-bound nature of the eigenstates is that an electron prepared in a state entirely within the well would, with a finite rate, tunnel through the potential barrier into the outside scattering region.  In the next section, we explore how to calculate this tunnelling rate, or equivalently the lifetimes of the quasi-bound states.

\section{Quasi-bound state lifetimes}
In this section, we investigate the lifetime of a quasi-bound electron within the potential well. We apply three different techniques:\ the WKB approximation, the finite difference method, and a phase shift method.

\subsection{Lifetimes from the WKB approximation}
\begin{figure}
    \centering
         \begin{subfigure}{0.49\linewidth}
        \includegraphics[width=\linewidth]{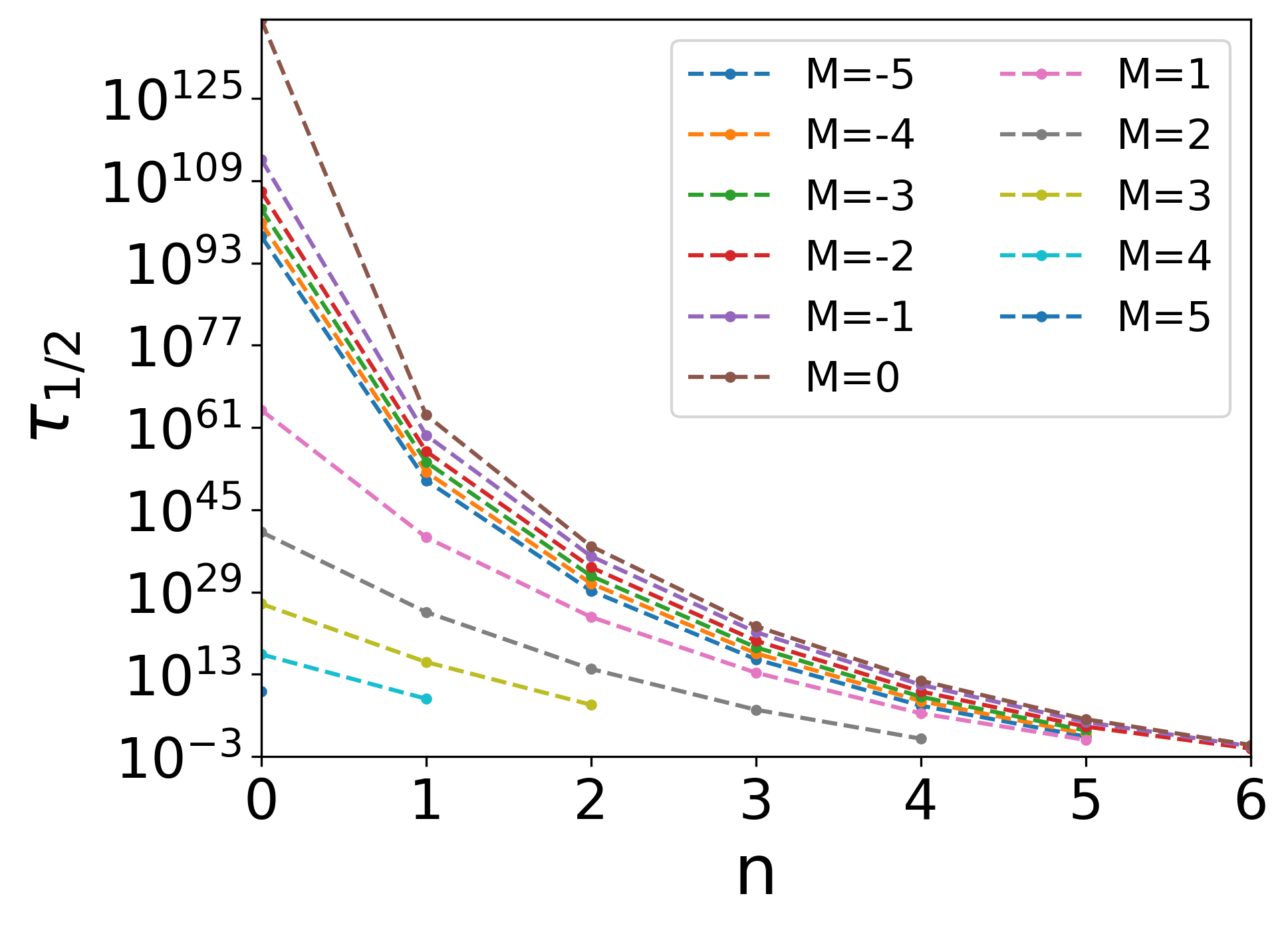}
        \caption{}
    \label{fig WKB lifetimes}
     \end{subfigure}
     \begin{subfigure}{0.49\linewidth}
         \includegraphics[width=\linewidth]{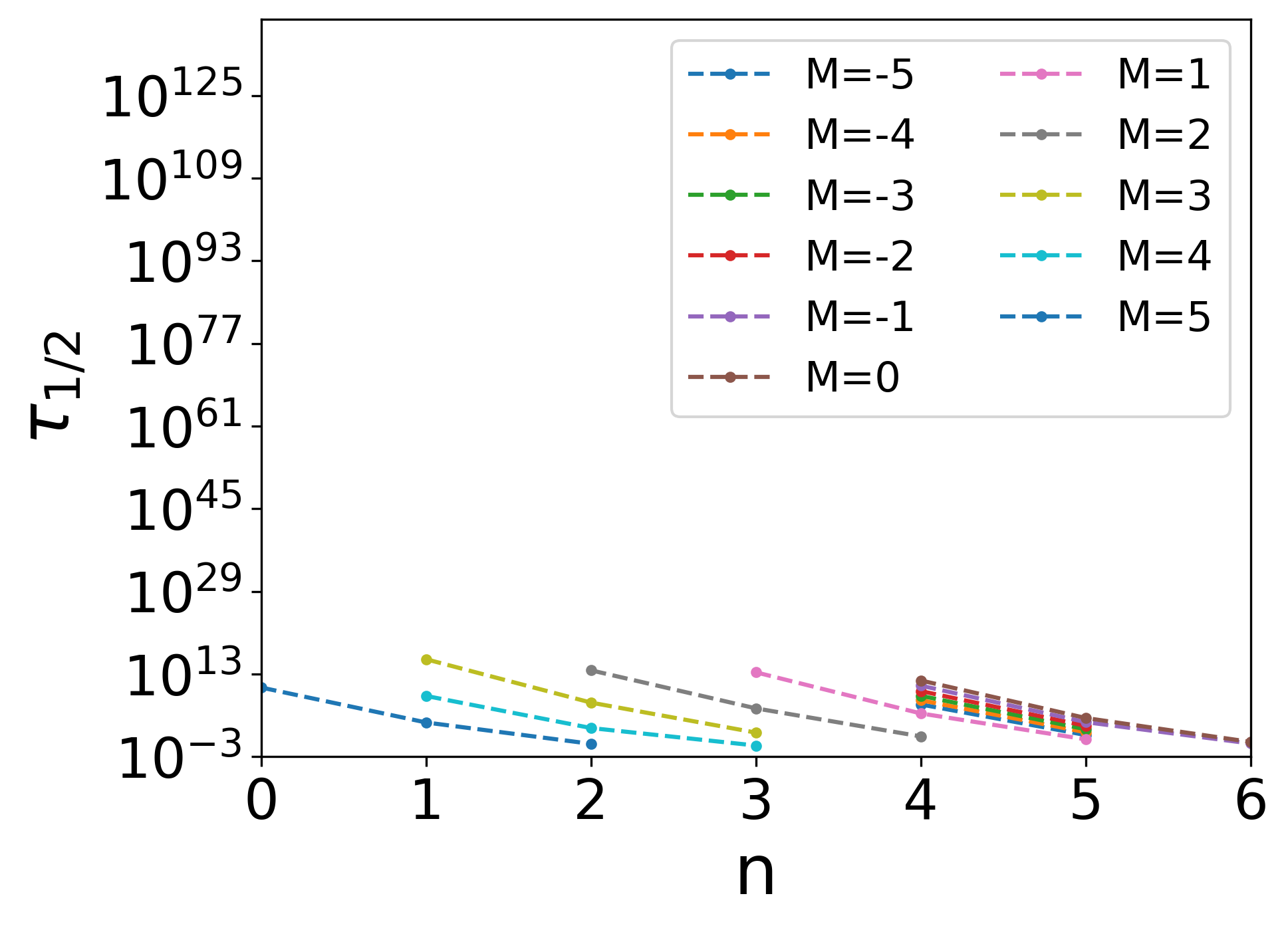}
         \caption{}
         \label{fig phase lifetimes}
     \end{subfigure}
    \caption{(a) The decay time \(\tau_{1/2}\) against the quantum number of the quasi-bound state \(n\) with $\lambda=100$ and different values of $M$, obtained via the WKB approximation.  (b) The decay time \(\tau_{1/2}\) against the quantum number of the quasi-bound state \(n\) with $\lambda=100$ and different values of $M$, obtained via  the phase shift method. The two methods predict similar half-lives.}
\end{figure}

The lifetime of a quasi-bound state of the electron can be modelled by considering it to be a particle bouncing back and forth within the well. Each time the particle meets a confining wall, there is a probability that it will tunnel through, and each successive bounce adds to the cumulative tunnelling probability.
The half-life, \(\tau_{1/2}\), may be estimated as
\begin{equation}
	\centering
	\tau_{1/2} \approx \frac{(\rho_2-\rho_1) \ln 2}{\sqrt{\epsilon_n}}\exp\left(2\int\limits_{\rho_2}^{\rho_3}P_n(\rho)d\rho\right).
\end{equation}
The derivation of this expression for \(\tau_{1/2}\) can be found in Appendix \ref{Appendix WKB lifetime}. The results of the lifetimes for \(\lambda=100\) and $M=-5$ to $5$ can be seen in Fig.~\ref{fig WKB lifetimes}.

\subsection{Lifetimes from the finite difference method}
\begin{figure}
    \centering
    \includegraphics[width=0.8\linewidth]{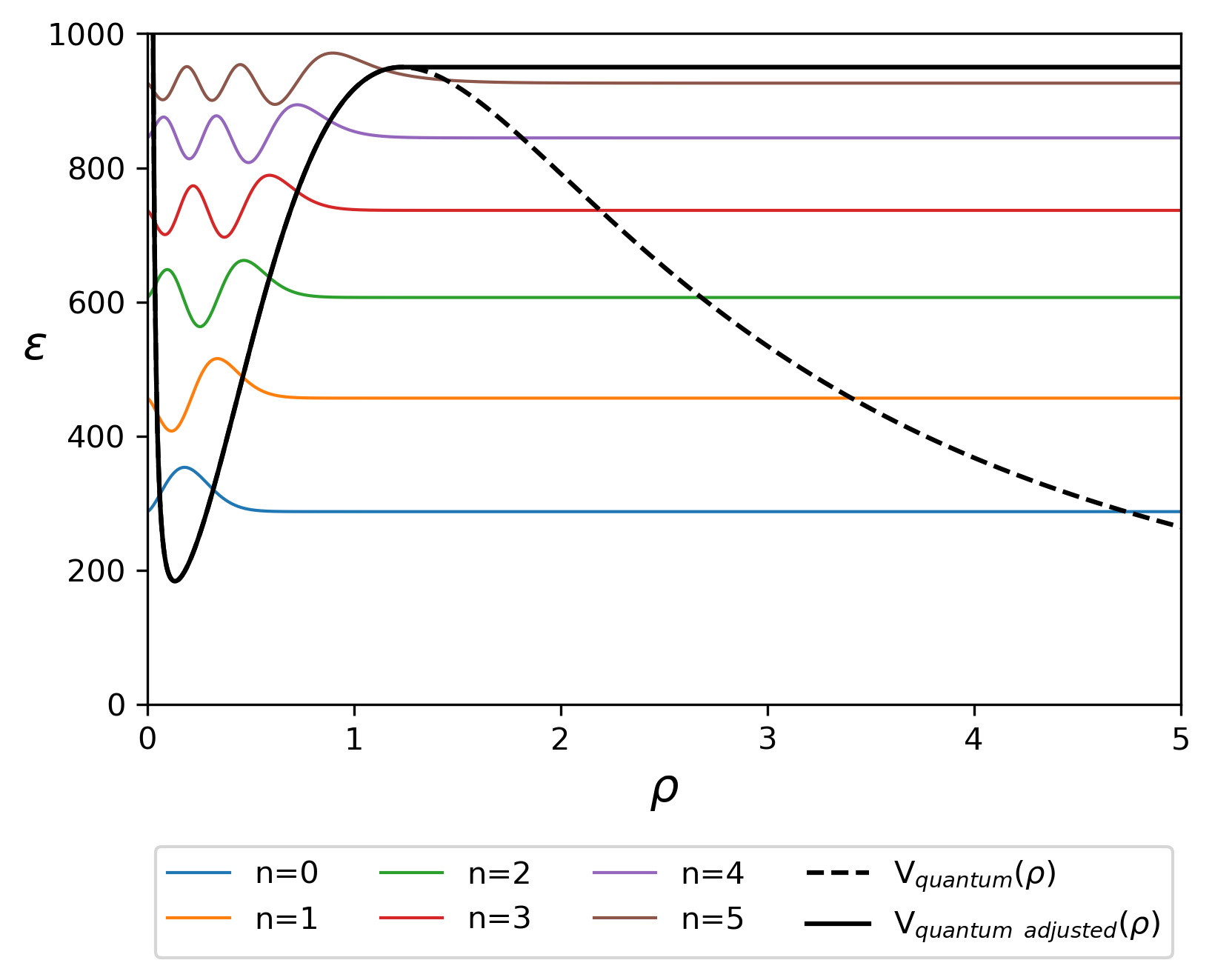}
    \caption{The original quantum potential, \(V_{\rm quantum}(\rho)\) (dashed black line), and its modified version, \(V_\mathrm{quantum~adjusted}(\rho)\) (solid black line). The eigenfunctions of the modified potential, \(\zeta_n(\rho)\), which are plotted offset by their energy eigenvalues, are now fully bound. This example is for the values \(M=1\) and \(\lambda=100\).}
    \label{fig adjusted potential}
\end{figure}
\begin{figure}
	\centering
	\includegraphics[width=0.8\textwidth]{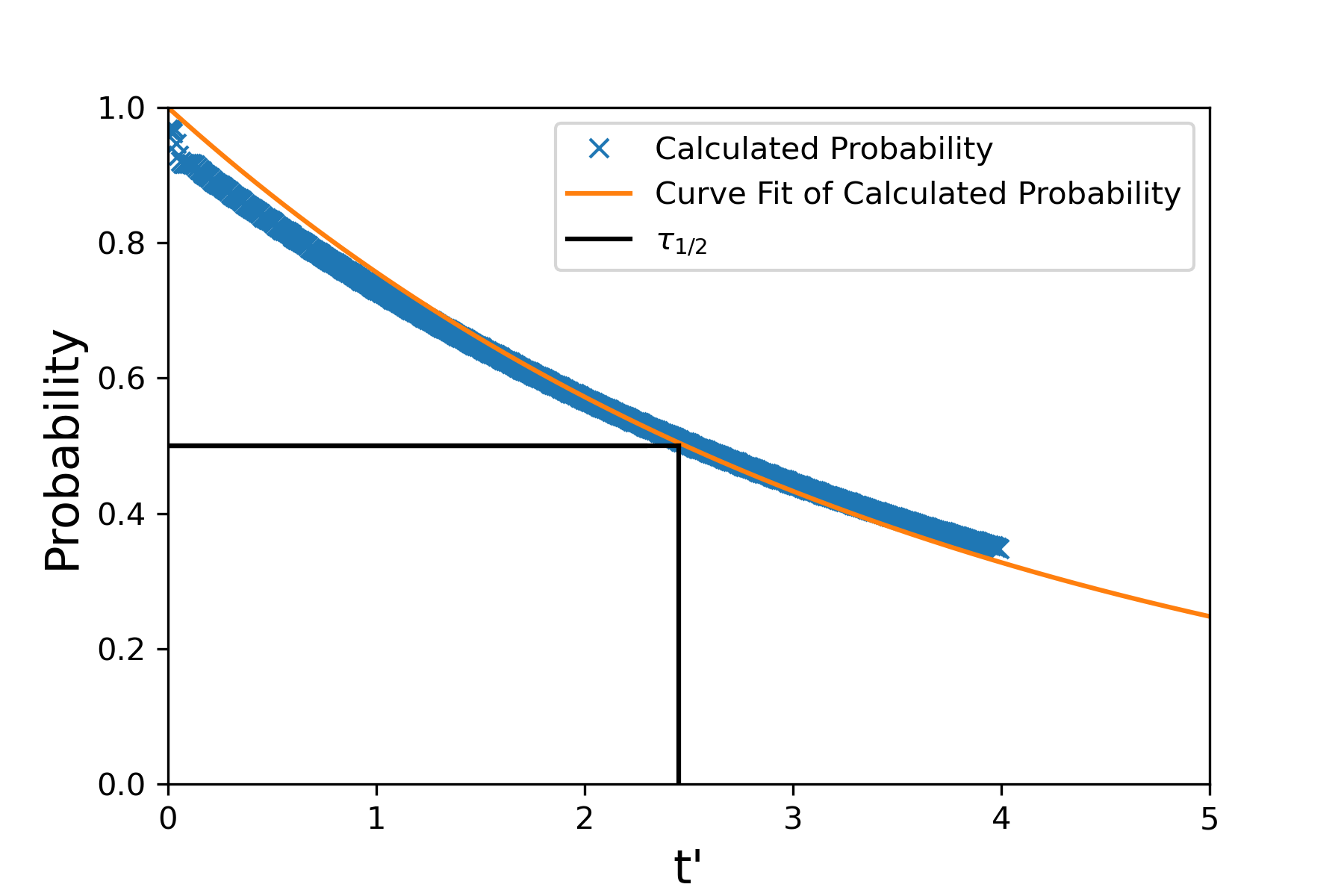}
	\caption{The estimated probability of the electron being found within the potential well as a function of the waiting time $t'$ (blue crosses).  The parameters are \(M=1\), \(\lambda=100\), and \(n=5\), and the estimates are made using the finite difference method with \(h=10000\) and \(\rho \in (0,160)\). The orange curve is an exponential fit to these results, \(P\approx \exp(-0.279\,t')\).  The half-life shown in this plot is \(\tau_{1/2}\approx2.45\).}
	\label{fig PLot of extended curve fitting}
\end{figure}

An alternative method for calculating the lifetime involves considering a modified potential that is constant beyond the position of the peak of the potential barrier:\ see Fig.~\ref{fig adjusted potential}.  A finite difference analysis then produces fully bound states with eigenfunctions \(\zeta_n(\rho)\), as shown in the figure.  These wave functions are at the same energies as the original quasi-bound wave functions.  We can then expand the fully bound wave function \(\zeta_n(\rho)\) in the basis of the eigenfunctions \(\psi_m(\rho)\) of our original problem, Eq. \eqref{Eq Quantum one dimensional schroinger equation}. Assuming that at \(t=0\) our electron in the quasi-bound problem is in the state with wave function $\zeta_n(\rho)$, the time-evolution of its wave function is approximately
\begin{equation}
    \centering
    \Psi(\rho, t')=\sum_{m=0}^N C_m\psi_m(\rho)e^{-i\epsilon_m  t'},~~~~\text{where}~~~~C_m=\left<\psi_m(\rho)\vert \zeta_n(\rho) \right>.
\end{equation}
An estimate for the half-life of the quasi-bound state can then be obtained by calculating the probability that the electron is still within the well and determining when this reaches 50 per cent.  The results of this procedure are shown in Fig.~\ref{fig PLot of extended curve fitting}. 

\subsection{Lifetimes from the phase shift method}
\begin{figure}
 	\centering
 	\includegraphics[width=0.8\textwidth]{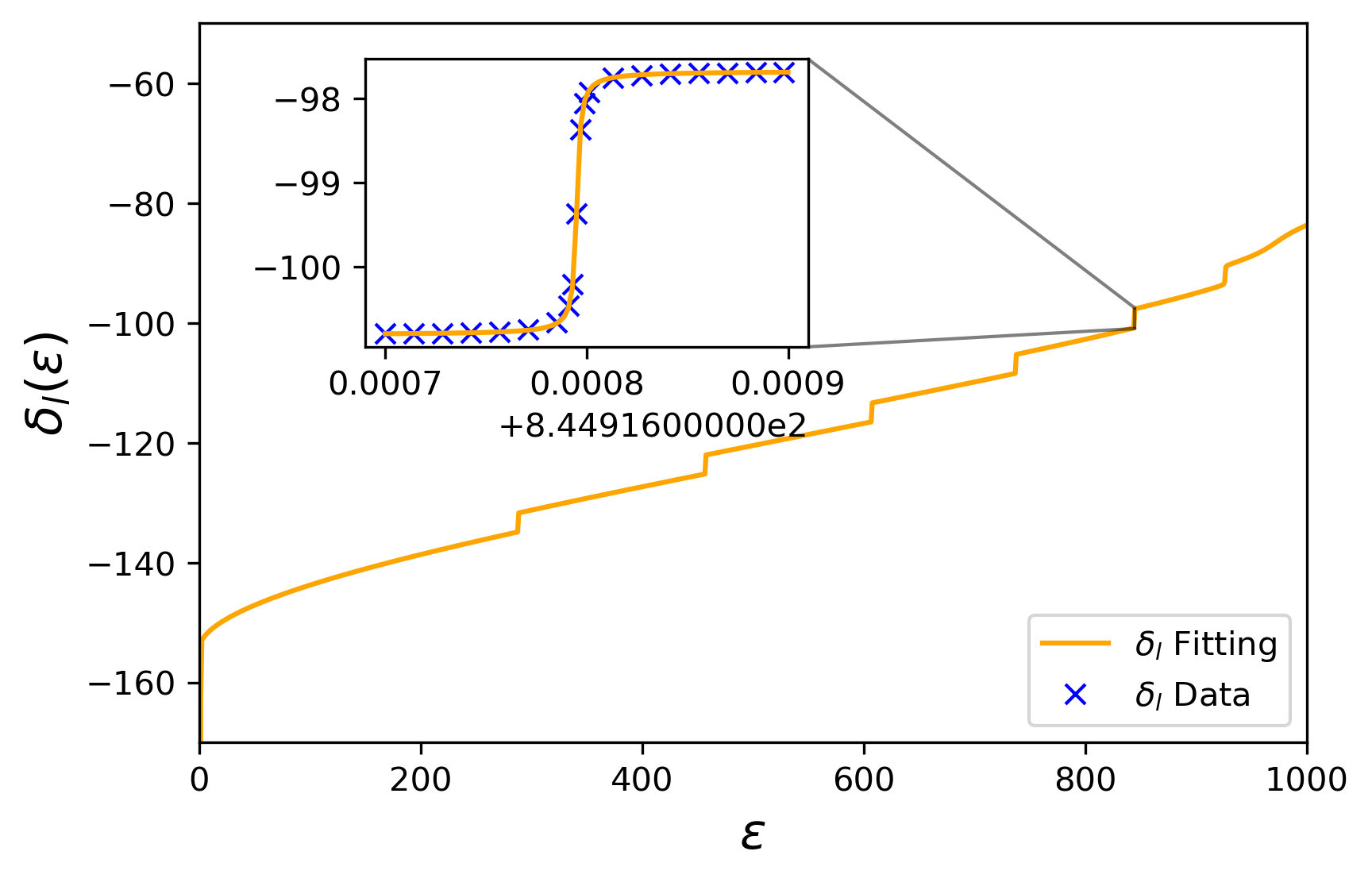}
 	\caption{The phase shift \(\delta_l(\epsilon)\) for \(M=1\) and \(\lambda=100\). Inset:\ A detailed view of the jump in \(\delta_l(\epsilon)\).  The data for this jump (blue crosses) are fitted with \(\delta_l(\epsilon)=\delta_l(\epsilon_n)+\arctan\left((\epsilon-\epsilon_n)/\Gamma\right)\) (orange curve), where \(\epsilon=844.91\), \(\Gamma=1.40\times10^{-6}\), and \(\delta_l(\epsilon_n)=-99.24\). }
 	\label{phase shift}
 \end{figure}

Consider the radial equation, (\ref{Eq dimensionless radial schrodinger equation}), in the absence of a magnetic monopole \((\lambda=0)\),
\begin{equation}
    -\frac{d^2G(\rho)}{d\rho^2} -\frac{1}{\rho}\frac{dG(\rho)}{d\rho}+\frac{M^2}{\rho^2}G(\rho)=\epsilon\, G(\rho).
\end{equation}
After changing the independent variable from \(\rho\) to \(z=\sqrt{\epsilon}\,\rho\), we get the Bessel equation,
\begin{equation}
    z^2\frac{d^2G(z)}{dz^2}+z\frac{dG(z)}{dz}+\left(z^2-M^2\right)G(z)=0.
    \label{eq bessel equation}
\end{equation}
The analytic solution of this equation is the Bessel function \(J_M(z)\), with asymptotes
\begin{equation}
    J_M(z)\sim
    \begin{cases}
     z^{|M|}, \qquad & z\rightarrow0, \\
         J_M(z)\sim\sqrt{\frac{2}{\pi z}}\cos{\left(z-|M|\frac{\pi}{2}-\frac{\pi}{4}\right)}, \qquad & z \gg 1.
  \end{cases}       
    \label{eq asymptotic behaviour}
\end{equation}
This suggests that we should write \(G(z)=\psi(z)/\sqrt{z}\), whereupon (\ref{eq bessel equation}) becomes
\begin{equation}
    -\frac{d^2\psi(z)}{dz^2}+\frac{M^2-1/4}{z^2} \psi(z) =\psi(z).
\end{equation}
Defining
\begin{equation}
    l=|M|-\frac{1}{2},
\end{equation}
the centrifugal potential \(\left(M^2-1/4\right)/z^2\) assumes the familiar form \(l\left(l+1\right)/z^2\) and the asymptotic behaviour (\ref{eq asymptotic behaviour}) for large $z$ becomes
\begin{equation}
    \psi(z)\sim \sqrt{\frac{2}{\pi}} \sin{\left(z-l\frac{\pi}{2}\right)}.
    \label{eq sayptote behaviour}
\end{equation}
Let us now compare this to the case when a magnetic monopole is present, i.e.\ when $\lambda \ne 0$. The radial equation (\ref{Eq dimensionless radial schrodinger equation}) becomes in terms of the new variable $z$ and the function \(\psi(z)=G(z)\sqrt{z}\), as
\begin{equation}
    -\frac{d^2\psi(z)}{dz^2}+V_\text{eff}(z,\epsilon)\psi(z)=\psi(z),
    \label{eq radial equation phase shift}
\end{equation}
where
\begin{equation}
    V_\text{eff}(z,\epsilon)=\frac{1}{z^2}\left\{-\frac{1}{4}+\left[M+\lambda\left(1-\frac{1}{\sqrt{1+z^2/\epsilon}}\right)\right]^2\right\}.
\end{equation}
The wavefunction's asymptotic behaviour turns out to be
\begin{equation}
    \psi(z)\sim \sin\left(z-l\frac{\pi}{2}+\delta_l(\epsilon, \lambda)\right),
    \label{eq asymptotic behaviour 1}
\end{equation}
where \(\delta_l(\epsilon,\lambda)\) is the phase shift, which depends on both the energy and the monopole strength \(\lambda\).

The characteristic energy-dependence of the phase shift around a resonance (quasi-bound state energy), jumping by \(\pi\) within a short range, can be seen in Fig.~\ref{phase shift}. Around each resonance, we can fit \(\delta_l(\epsilon)\) by
\begin{equation}
    \delta_l(\epsilon)=\delta_l(\epsilon_n)+\arctan\left(\frac{\epsilon-\epsilon_n}{\Gamma}\right),
\end{equation}
where \(\epsilon_n\) is the dimensionless energy of the resonance. Such an energy dependence of the phase shift implies a time delay
\begin{equation}
    \tau\sim\frac{1}{2\Gamma}
\end{equation}
between the arrival of an incoming wave-packet and the emission of an outgoing scattered wave-packet. This time delay is an estimate of the quasi-bound state lifetime.

The phase shift as a function of the energy was computed using the variable phase method of Morse and Allis \cite{RefWorks:18}, which is based on writing the solution of the radial equation (\ref{eq radial equation phase shift}) as 
\begin{equation}
    \psi(z)=C(z)\sin\left(z+\mu(z)\right),
    \label{eq asymptoitc phase shift}
\end{equation}
where the variable phase \(\mu(z)\) and the variable amplitude \(C(z)\) must be suitably related. Comparing (\ref{eq asymptotic behaviour 1}) and (\ref{eq asymptoitc phase shift}), we have
\begin{equation}
    \delta_l(\epsilon)=\lim_{z\rightarrow\infty}\mu(z)-l\frac{\pi}{2}.
    \label{Eq delta l phase shift}
\end{equation}
As described in Appendix \ref{Appendix muz derivation}, \(\mu(z)\) satisfies
\begin{equation}
    \frac{\partial \mu(z)}{\partial z}=V_\text{eff}(z)\sin^2\left(z+\mu(z)\right),
\end{equation}
which can be solved numerically from $z=0$ up to a large enough $z$ to allow an accurate computation of the phase shift \(\delta_l(\epsilon)\).

\subsection{Comparison of lifetime results from different methods}

The three different methods to calculate lifetimes above all involve different approximations and thus have different regimes of applicability.  
\begin{itemize}
\item The WKB method is by far the easiest to use, but is a semi-classical method.  This makes it most reliable at higher quantum numbers, although it also turns out that it is the only method that can be practically applied for the longer lifetimes at low quantum numbers.
\item The finite difference method, while seemingly approximation free as the exact Schr\"odinger equation is solved numerically, is limited by practical computation power.  First, the finite size $R_\text{max}$ of the numerical box means that it suffers from self-interference due to the reflection of the outgoing wave at the system boundary before a half-life can be evaluated, meaning that very long lifetimes cannot be computed.  This can be overcome by using an increased spatial range required and a  smaller step size in the finite difference method, but at the cost of more computational power, which scales like $R_\text{max}^3$ assuming the grid discretisation is kept fixed.

Conversely, for states near the top of the well with short lifetimes, the significant penetration of the wavefunction into the barrier leads to ambiguity in defining which part of the wavefunction is bound.  This in turn leads to ambiguity of whether the electron is in the well or not, meaning the choice of convention associated with assigning any definite lifetime may not be exactly the same as in the other methods.
\item The phase shift method requires numerical integration of an ODE \eqref{eq:RKeq} which is performed using a 4th order Runge-Kutta method with completely controlled accumulated numerical error.  In this sense, the phase shift method may be considered the most reliable, however for large lifetimes, the width of the jump becomes very small and can not be measured when this width reaches machine precision.
\end{itemize}

\begin{figure}
    \centering
    \includegraphics[width=0.8\linewidth]{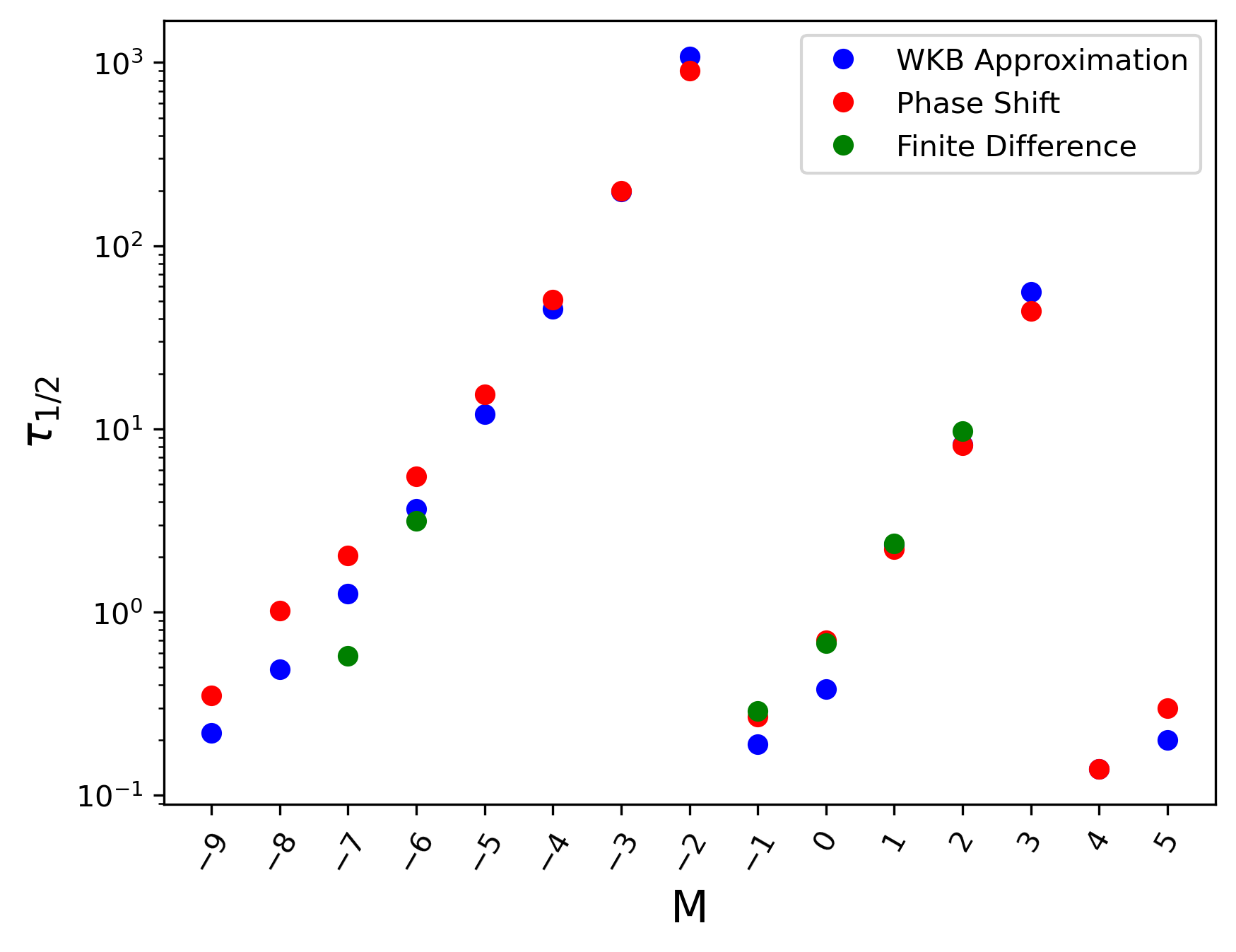}
    \caption{The highest-energy quasi-bound state half-lives \((\tau_{1/2})\) for \(-9 \leqslant M \leqslant 5\), where \(\lambda=100\). Good agreement can be seen between the results of all three calculation methods:\ the WKB approximation, the phase shift method, and the finite difference method.}
    \label{Fig eigenvalues lifetime for full range of m}
\end{figure}

In Fig.~\ref{Fig eigenvalues lifetime for full range of m}, we compare the estimates of the lifetimes obtained with the three different methods, considering the highest energy quasi-bound states for several values of \(M\) and \(\lambda=100\).  In the lifetime ranges where they are applicable (which are the only ones shown) we can see good agreement between all three methods of calculating the quasi-bound-state lifetime: the phase shift method, the WKB approximation method, and the finite difference method.
 
\section{Discussion}
\label{sec:discussion}

Before discussing the application of our results to real experimental systems, let us return to the previous work that considered electrons on a plane in the vicinity of a monopole, Ref.~\cite{Diamantini2022}.  As the motivation of that paper is a pairing mechanism for superconductivity it may naively appear unrelated, however within the approximations made in that work in order to make the problem analytically tractable, it reduces in essence to the one we look at here of independent electrons in planes above/below a monopole.  One obvious difference between their work and ours is that they include the Zeeman term, which will create an extra contribution to the potential in the form of a potential well centred at the origin.  However more importantly, they neglect the ${\bf A}^2$ term in the kinetic part of the Hamiltonian, on grounds that do not apply in the full case that we consider here.  It would certainly be interesting future work to apply our detailed results of this single electron problem to the case of pairing.

As discussed in the introduction, even though no actual Dirac monopoles have been discovered in nature, close approximations to them can be found. Five sources of magnetic monopole analogues are magnetic needles, spin ice, artificial spin ice, magnetoelectric materials, and Josephson junction arrays.  Were we to place a 2D electron gas in the vicinity of one of these systems, what would the requirements be for the formation of a quasi-bound state due to the monopole-like field emanating from the system?

%\subsection{Expected half-life}
As shown above, the minimum monopole charge required for a single quasi-bound state to appear in the 2DEG's electronic spectrum is 18\(Q_D\), which results in a quasi-bound state with angular momentum quantum number $M=0$. In this case the dimensionless half-life is 
%\(\tau_{1/2}=1.16\) (finite difference method), or 
\(\tau_{1/2}\approx 4\) (phase shift method).  The associated dimensionful half-life (see Eq.~\eqref{eq:t0}) is given by
\begin{equation}
    t(D)=\frac{2\tau_{1/2}m^{*}D^2}{\hbar}.
\end{equation}
Using this equation and the distances between monopoles in the various monopole analogues we can ascertain the approximate lifetimes that one might expect
%. Using both the shorter half-lives from the finite difference method and the longer ones from the phase shift method we can get a range of predicted half-lives 
in SI units.

\subsection{Magnetic Needle}
 Current realisations of magnetic needles achieve \(Q_m\) of order \(6Q_D\) \cite{RefWorks:17}; this is approximately one third of the magnetic monopole charge required to produce a quasi-bound state. Let us model the tip of the needle as a sphere with a radius of \(100\,{\rm nm}\). Assuming that the magnetic field strength at the surface of the sphere equals
the magnetic field applied to the physical needle to magnetise it, which is of order $0.15\,{\rm T}$, we can estimate the monopole charge at the needle tip from
\begin{equation}
    \centering
    Q_m=\frac{4 B \pi r^2}{\mu_0}.
\end{equation}
We obtain \(Q_m \approx 4.6Q_D\) (see Appendix~\ref{Appendix magnetic needle}), which is in agreement
with the observed value. If the magnetic field strength at the tip of the needle can reach a value of $\sim 0.6\,{\rm T}$ then the magnetic charge of the needle tip will be $\sim 18Q_D$, the required minimum to find a quasi-bound state. The aperture radius around the needle tip is approximately \(10\,\mu{\rm m}\) \cite{RefWorks:17}, so if we assume the electron is found at half this distance then the approximate half-life is of order of $1\,\mu{\rm s}$. 

\subsection{Spin-Ice}
The magnetic charge of a monopole in spin ice is given by \cite{Castelnovo2008}:
\begin{equation}
    Q_m=\frac{\mu}{\mu_B}\frac{\alpha\lambda_C}{\pi a_d}Q_D \approx\frac{Q_D}{8000},
\end{equation}
where \(\alpha\) is the fine structure constant, \(\mu_B\) is the Bohr magneton, \(\lambda_C\) is the Compton wavelength for an electron, \(\mu\) is the magnetic permeability, and \(a_d\)
is the distance between the centres of neighbouring tetrahedra of rare-earth sites in the pyrochlore crystal structure. The magnetic charge of a monopole in spin ice would need to be 144000 times larger to achieve the threshold of \(18Q_D\) required for the formation of a single quasi-bound state in the 2DEG. In ref.~\cite{Castelnovo2008} the authors state that the charge of the monopole can be tuned by changing the pressure applied to the crystal, causing changes in the value of \(\mu/a_d\). If a pressure could be achieved such that
\begin{equation}
    \frac{\mu}{a_d}= \frac{18\mu_b\pi}{\alpha\lambda_C},
\end{equation}
then a spin-ice monopole would become sufficient to form a quasi-bound state for an electron of \(M=0\). The pressures needed are so great that realistically this could not be achieved.

\subsection{Artificial Spin-Ice}
The magnetic charge at any given lattice site is given by \cite{Ladak2010,Chern2011}
\begin{equation}
    \centering
    Q_\alpha=\sum_i q_i,
\end{equation}
where \(Q_\alpha\) is the magnetic charge at the monopole site and \(q_i\) is the charge corresponding to one end of a magnetic dipole whose length is the lattice spacing. As a concrete example, consider a square lattice of bar magnets, so that each lattice site has an end of one of four bar magnets sitting on it \cite{Moller2006}, all of which have the same magnitude of the charge \(|q_i|=q\). \(Q_\alpha\) can then take one of five values:\ $-4q$, $-2q$, $0$, $2q$, or $4q$. To reach a value where \(Q_\alpha=18Q_D\) would require \(q=9Q_D\) for the \(Q_\alpha=\pm2q\) states or \(q=9Q_D/2\) for the \(Q_\alpha=\pm4q\) states. The distances given in refs.~\cite{Chern2011,Moller2006} range from \(100\) to $1000\,{\rm nm}$, so the lifetimes of the quasi-bound state associated with the minimum necessary monopole charge in artificial spin ice are in the range $0.7$ to $70\,{\rm ns}$.

\subsection{Image charges in magnetoelectric materials}

In the prototypical magnetoelectric material Cr${}_2$O${}_3$, an impurity charge of one electron at the surface induces a monopole charge of order $10^{-16}$Am \cite{Meier2019}.  This is eight orders of magnitude smaller than the threshold of $18Q_D$ required to induce at least one quasi-bound state in the 2DEG.  While larger impurity charges would give proportionally larger monopole charges, finding a regime where the force on the 2DEG is dominated by the induced monopole and not the impurity charge seems unlikely.

On the other hand, the equivalent topological magnetoelectric effect which can be realised by depositing a magnetic layer on the surface of a three-dimensional topological insulator such as Bi${}_{1-x}$Sb${}_x$ is much larger \cite{Qi2009,Uri2020}.  Here, a single electron impurity charge naturally induces a Dirac monopole at the surface.  As a layered nanostructure, the 2DEG could then in principle be deposited on top of the magnetic layer at a distance \(10\) to $100\,{\rm nm}$, thus giving lifetimes in the range of \(7\) to $700\,{\rm ps}$.  Further calculation would be required however in order to optimise the setup so the monopole potential dominates in the 2DEG and not that from the impurity charge.

\subsection{Emergent quantum excitations in Josephson junction arrays and superconducting films.}

Like spin-ice, granular superconductors \cite{Diamantini2021,Diamantini2025} and a nano-structured counterpart, Josephson Junction arrays \cite{Trugenberger2020}, are predicted to have magnetic monopole excitations.  Unlike spin-ice, these are predicted to have quantised Dirac charge.  While individual monopole excitations have to date not been experimentally measured, their existence can be inferred from the super-insulating state which can be interpreted as a Bose condensate of magnetic monopoles.   In the region between the superconductor and the superinsulator, one may expect a low density of (relatively) isolated monopole excitations to exist in these systems, which would allow the physics described in this paper to be probed. 
At present however, there are no appropriate experimental measurements of the properties of such isolated monopoles that would allow us to make more definite predictions.  This is a promising and fertile experimental area for further exploration.

\section{Conclusion}
In this paper we have demonstrated that a magnetic monopole can create bound states (in the classical case) or quasi-bound states (in the quantum one) for a charged particle confined to a plane adjacent to the monopole.

Classically, the orbits are either bound or scattered. Depending
on the monopole strength and the energy and angular momentum of the electron,
the classically bound orbits fall into four classes. All classical solutions have at least one circular orbit; all other orbit types are as shown in Fig.~\ref{Fig classical all classical orbits}. 

Computing the eigenenergies in the quantum problem, we found a good agreement between the Bohr-Sommerfeld approach and Schr{\"o}dinger-equation quantum mechanics. If we replace the classical potential with the quantum potential, the Bohr-Sommerfeld quantisation procedure produces the same eigenvalues as the WKB approximation (see Fig.~\ref{Fig Eigenvalues bohr with quantum potential}). The difference between the classical potential and the quantum one is a term \(-1/4\rho^2\) which arises from a transformation of the variables in the Laplacian.

In the quantum solution there are no true bound states, but we find quasi-bound states with a finite lifetime, which can be large. Larger values of monopole charge are required to produce a single quasi-bound state for positive values of angular momentum than for negative values (see Fig.~\ref{Fig min lamba and qm}). We have made estimates for the lifetimes of the quasi-bound states using three different methods (WKB, finite difference, and phase-shift). All three methods give roughly similar results, although they each have their own regimes of applicability (see Fig.~\ref{Fig eigenvalues lifetime for full range of m}).

In the absence of real Dirac monopoles, one of tfive options --- spin ice, artificial spin ice, granular superconductors, magnetoelectric surfaces and nanoscale magnetic needles --- could be used to simulate them. We have investigated whether any of these systems can meet the minimum
value of the magnetic monopole charge required to form at least one quasi-bound state. Currently reported values for monopole charges simulated with nanoscale magnetic needles, artificial spin-ice systems, or topological insulators are below the threshold, though not impossibly far below it. 
%If the threshold were achieved, a quasi-bound state with lifetime of the order of $\mu{\rm s}$ would be expected using nanoscale magnetic needles, while much smaller lifetimes of the order of $0.2$ to $20\,{\rm ns}$ would be expected using artificial spin ices.  
For solid-state spin ice crystals or non-topological magnetoelectric surfaces, however, the current reported values of simulated monopole charges are much smaller than the threshold, which therefore seems to be unattainable in such systems.  Granular superconductors may meet the threshold, but require more experimental work to isolate the parameter regime where monopole excitations are dilute.

Finally, in all our calculations we neglected the electron spin.  This allowed us to fully analyse the physics arising from the orbital interaction of the electron and the magnetic field of the monopole, including the curious result that the distance to the monopole can be scaled out completely, even in the quantum problem.  A monopole very far away from the two-dimensional surface can still create a (quasi-)bound state, however the binding energy will be very small.  In contrast, the strength of the monopole can be scaled out in the classical problem, but not the quantum one.  In an experimental situation, this could be probed by using a Cooper pair (within a superconductor) as a probe rather than a single electron, or for a two-dimensional electron gas material with a small or vanishing $g$-factor.  In any other setup, the effect of the spin would have to be added to the theoretical model for the particular geometry under consideration.

\section{Acknowledgments}
J. Quintanilla acknowledges support from the EPSRC under the project `Unconventional superconductors: new paradigms for new materials' (Grant No. EP/P00749X/1) and thanks Claudio Castelnovo and Stephen Blundell for stimulating discussions.

C. Hooley is grateful for financial support from UKRI via grant number EP/R031924/1. This work was performed in part at Aspen Center for Physics, which is supported by National Science Foundation grant PHY-2210452.

G. M\"oller acknowledges support from the Royal Society under a University Research Fellowship (grants UF120157 and URF\textbackslash R\textbackslash 180004); G. M\"oller and J. Pooley were supported by Royal Society Enhancement Award RGF\textbackslash EA\textbackslash 180107.
%###############end of main body of text#############

\appendix
\setcounter{equation}{0}
\renewcommand{\theequation}{\thesection.\arabic{equation}}
%#################appndix########################
\section{The vector potential of a magnetic monopole}\label{Appendix vector potential}
The vector potential for a magnetic monopole can be written in the form of two overlapping non-singular potentials. It is shown in refs.~\cite{RefWorks:33,RefWorks:32} that these vector potentials can be expressed as
\begin{equation}
\centering
\mathbf{A}=
\left\{
\begin{array}{ccc}
    \displaystyle \frac{\mu_0 Q_m}{4 \pi R}\frac{\left(1-\cos{\theta}\right)}{\sin{\theta}}\hat{\mbox{\boldmath $\phi$}},
    & \qquad & \displaystyle \theta \leqslant \frac{\pi}{2}; \\
    & & \\
    \displaystyle \frac{\mu_0 Q_m}{4 \pi R}\frac{\left(-1-\cos{\theta}\right)}{ \sin{\theta}}\hat{\mbox{\boldmath $\phi$}}, & & \displaystyle \theta \geqslant \frac{\pi}{2} .
    \end{array} \right.
\end{equation}
Since in our case the plane is taken to lie above the monopole, the required expression for \(\mathbf{A}\) is the first of these.  From Fig.~\ref{fig graphical representation of the system}
we see that
\begin{equation}
R=\sqrt{r^2+D^2},
\qquad 
\cos{\theta}=\frac{D}{\sqrt{r^2+D^2}},
\qquad
\mbox{and}
\qquad
\sin{\theta}=\frac{r}{\sqrt{r^2 +D^2}};
\end{equation}
hence \(\mathbf{A}\) can be expressed in terms of $r$  as
\begin{equation}
    \centering
    \mathbf{A}(r)=\frac{\mu_0 Q_m}{4 \pi r}\left(1-\frac{D}{\sqrt{r^2+D^2}}\right)\hat{\mbox{\boldmath $\phi$}}.
\end{equation}

\section{Semiclassical estimate of the number of quantum bound states}\label{appendix derivation of N}
Consider a well in the effective potential $V_{\rm cl}/\lambda^2$ of width $w_{1/2}$ and depth $\epsilon_{1/2}$.  This corresponds to a dimensionful width of $\delta x = w_{1/2} D$ and a dimensionful depth of $\delta E = E_0 \lambda^2 \epsilon_{1/2}$.

Let us approximate this potential well using the harmonic oscillator form $V = \kappa x^2 / 2$.  This allows us to estimate the spring constant $\kappa$ for the well in question:
\begin{equation}
\kappa \approx \frac{\delta E}{(\delta x)^2} = \frac{E_0 \lambda^2 \epsilon_{1/2}}{w_{1/2}^2 D^2}. 
\end{equation}
The number of bound states contained in such a well below energy $\delta E$ is given approximately by
\begin{equation}
N \approx \frac{\delta E}{\hbar \omega_0} = \frac{\delta E}{\hbar} \sqrt{\frac{m^{*}}{\kappa}} = \frac{E_0 \lambda^2 \epsilon_{1/2}}{\hbar} \sqrt{\frac{m^{*} w_{1/2}^2 D^2}{E_0 \lambda^2 \epsilon_{1/2}}} = \frac{\lambda D w_{1/2}}{\hbar} \sqrt{m^{*} E_0 \epsilon_{1/2}}.
\end{equation}
Using the definition of $E_0$ from (\ref{energyscale}), this can be simplified to
\begin{equation}
N \approx w_{1/2} \lambda \sqrt{\frac{\epsilon_{1/2}}{2}},
\end{equation}
the formula used in the main text.

\section{Derivation of the radial part of the Schr{\"o}dinger equation}\label{appendix radial equation}

We assume a solution of the form \(\Psi\left(r,\phi\right)=G(r)\,e^{iM\phi}\), where \(G(r)\) is a function of $r$, $M$ is the angular momentum quantum number, and \(A_\phi(r)\) is given by (\ref{Eq A dimesionfull}). Substituting this form into the Schr\"odinger equation (\ref{tise2d}), we obtain
\begin{equation}
    \centering
    \left(-\frac{\hbar^2}{2m^{*}}\frac{d^2}{d r^2}-\frac{\hbar^2}{2m^{*}r}\frac{d}{dr} + \frac{\hbar^2 M^2 }{2m^{*}r^2} -\frac{g \hbar q_e M  H(r)}{2 m^{*} D^2} + \frac{g^2 q_e^2 r^2 H(r)^2}{8 m^{*} D^4}\right)G(r) = E\,G(r),
    \label{Eq Quantum radial schrodinger equation}
\end{equation}
where
\begin{equation}
    \centering
    g=\frac{Q_m \mu_0}{4 \pi}=\frac{\lambda \hbar}{-q_e}, \qquad \qquad H(r)=\frac{2 D^2}{r^2}\left(1 - \frac{D}{\sqrt{D^2+r^2}}\right). \label{appdefs1}
\end{equation}
Using the definitions of $\rho$ and $\epsilon$ in terms of $r$ and $E$ given in (\ref{r_scale}) and (\ref{E_scale}) respectively, and (\ref{appdefs1}), this may be rewritten as
\begin{equation}
\left( - \frac{d^2}{d\rho^2} - \frac{1}{\rho} \frac{d}{d\rho} + \frac{M^2}{\rho^2} + \lambda M H + \frac{\lambda^2\rho^2}{4} H^2 \right) G = \epsilon\,G,
\end{equation}
or, collecting the non-derivative terms together,
\begin{equation}
\left( - \frac{d^2}{d\rho^2} - \frac{1}{\rho} \frac{d}{d\rho} + \frac{\lambda^2}{\rho^2} \left[ 
\frac{M}{\lambda} + \frac{\rho^2 H}{2} \right]^2 \right) G = \epsilon\,G,
\end{equation}
as claimed in the main text.

\section{Modelling a magnetic needle as a monopole}\label{Appendix magnetic needle}
Using (\ref{Eq maxwell monopole}), the Maxwell equation modified to account for magnetic monopoles, and assuming the tip of the needle is a sphere with a radius of \(100\,\text{nm}\), the magnetic charge of the needle is
\begin{equation}
    \centering
    Q_m=\frac{4B\pi r^2}{\mu_0}.
\end{equation}
Ref.~\cite{RefWorks:17} reports a $0.15\,{\rm T}$ magnetic field of the needle, so \(Q_m=1.5\times10^{-8}\text{Am}\), assuming the field of the magnetically charged needle matched the field that was applied. A unit of Dirac charge is given by (\ref{eq dirac charge}), which in SI units implies that \(Q_D=3.291\times10^{-9}\text{Am}\). This gives a resulting magnetic needle charge of \(\approx 4.6\,Q_D\).

\section{WKB lifetime}\label{Appendix WKB lifetime}
Every time the particle bounces against the wall of the potential well there is a probability, \(C_T\), that it will be able to tunnel out:
\begin{equation}
	\centering
	C_T = \exp(2J_n) = \exp\left(-2\int\limits_{\rho_2}^{\rho_3}P_n(\rho)\,d\rho\right).
\end{equation}
 Assuming that the particle moves ballistically within the well, the time $t'$ for a single bounce from the potential barrier and back again can be determined from the kinetic energy:
 \begin{equation}
     \centering
 \epsilon_n = \frac{\left(\rho_2-\rho_1\right)^2}{(t')^2}
 \qquad \longrightarrow \qquad
     t'=\frac{\left(\rho_2-\rho_1\right)}{\sqrt{\epsilon_n}}.
 \end{equation}
The time taken to perform enough tunnelling attempts to get through the barrier may be estimated as
 \begin{equation}
     \tau=t'/C_T,
 \end{equation} 
which therefore gives an expression for the half-life of
 \begin{equation}
     \tau_{1/2}=\tau\ln{2}.
 \end{equation}

\section{Variable phase method}\label{Appendix muz derivation}

The free electron radial solution is given as
\begin{equation}
    u(z) = C(z)\sin(z+\mu(z)).
    \label{Eq phase shift solution z}
\end{equation}
Differentiating (\ref{Eq phase shift solution z}) with respect to $z$ gives
\begin{equation}
	\centering
	\frac{d u(z)}{d z}=\frac{d C(z)}{d z}\sin\left(z+\mu(z)\right) + C(z) \left( 1 + \frac{d \mu(z) }{d z} \right) \cos \left( z + \mu(z) \right).
	\label{Eq First derivative of u(z)}
\end{equation}
The first derivative of \(u(z)\) is required to be \cite{RefWorks:18}
\begin{equation}
\frac{d u(z)}{d z}=C(z)\cos(z+\mu(z)),
\label{EQ first derivative pof mu required}
\end{equation}
which imposes the following relation between the variable amplitude and the variable phase:
\begin{equation}
	\centering
	\frac{d C(z)}{d z}\sin\left(z+\mu(z)\right) = -C(z)\frac{d \mu(z)}{d z}\cos\left(z+\mu(z)\right).
	\label{Eq equality in firs derivative of u(z)}
\end{equation}
Differentiating (\ref{EQ first derivative pof mu required})  with respect to $z$, we obtain
\begin{equation}
	\centering
	\frac{d^2 u(z)}{d z^2}= \frac{d C(z)}{d z}\cos \left(z+\mu(z)\right) -C(z) \left(1+\frac{d \mu(z)}{d z}\right) \sin \left(z + \mu(z)\right).
	\label{Eq second derivative of u(z)}
\end{equation}
Substituting \(\frac{d C(z)}{d z}\) from (\ref{Eq equality in firs derivative of u(z)}) into (\ref{Eq second derivative of u(z)}), we obtain
\begin{equation}
	\centering
	\frac{d ^2 u(z)}{d z^2} = - C(z) \frac{d \mu(z)}{d z} \frac{1}{\sin \left(z+\mu(z)\right)} - C(z)\sin(z+\mu(z)).
	\label{Eq appendix 2 second derivative of u}
\end{equation}
The Schr\"odinger equation in terms of $u(z)$ including a magnetic monopole is
\begin{equation}
    -\frac{d^2u(z)}{d z^2} +\left(V(z)-1\right)u(z)=0.
    \label{Eq phase shift schrodinger in z}
\end{equation}
Substituting (\ref{Eq appendix 2 second derivative of u}) and (\ref{Eq phase shift solution z}) into the Schr\"odinger equation (\ref{Eq phase shift schrodinger in z}) gives
\begin{equation}
	\centering
	C(z)\frac{d \mu(z)}{d z}\frac{1}{\sin (z +\mu(z))}+ C(z)\sin(z+\mu(z)) +(V(z)-1)C(z)\sin(z+\mu(z))=0,
\end{equation}
which simplifies to
\begin{equation}
	\centering
	C(z)\frac{d \mu(z)}{d z}\frac{1}{\sin (z +\mu(z))}+V(z)C(z)\sin(z+\mu(z))=0,
	\label{eq:RKeq}
\end{equation}
whence it follows that
\begin{eqnarray}
	\frac{d \mu(z)}{d z}&=&-V(z)\sin^2(z+\mu(z)) \\
	&=&-\left(-\frac{1}{4}+\left[M+\lambda\left(1-\frac{1}{\sqrt{1+\frac{z^2}{\lambda \epsilon}}}\right)\right]^2\right)\frac{\sin^2(z+\mu(z))}{z^2}.
\label{Eq diff mu}
\end{eqnarray}
This is the differential equation we need to solve numerically to compute the
phase shift from (\ref{Eq delta l phase shift}).

\begin{figure}
	\centering
	\includegraphics[width=\textwidth]{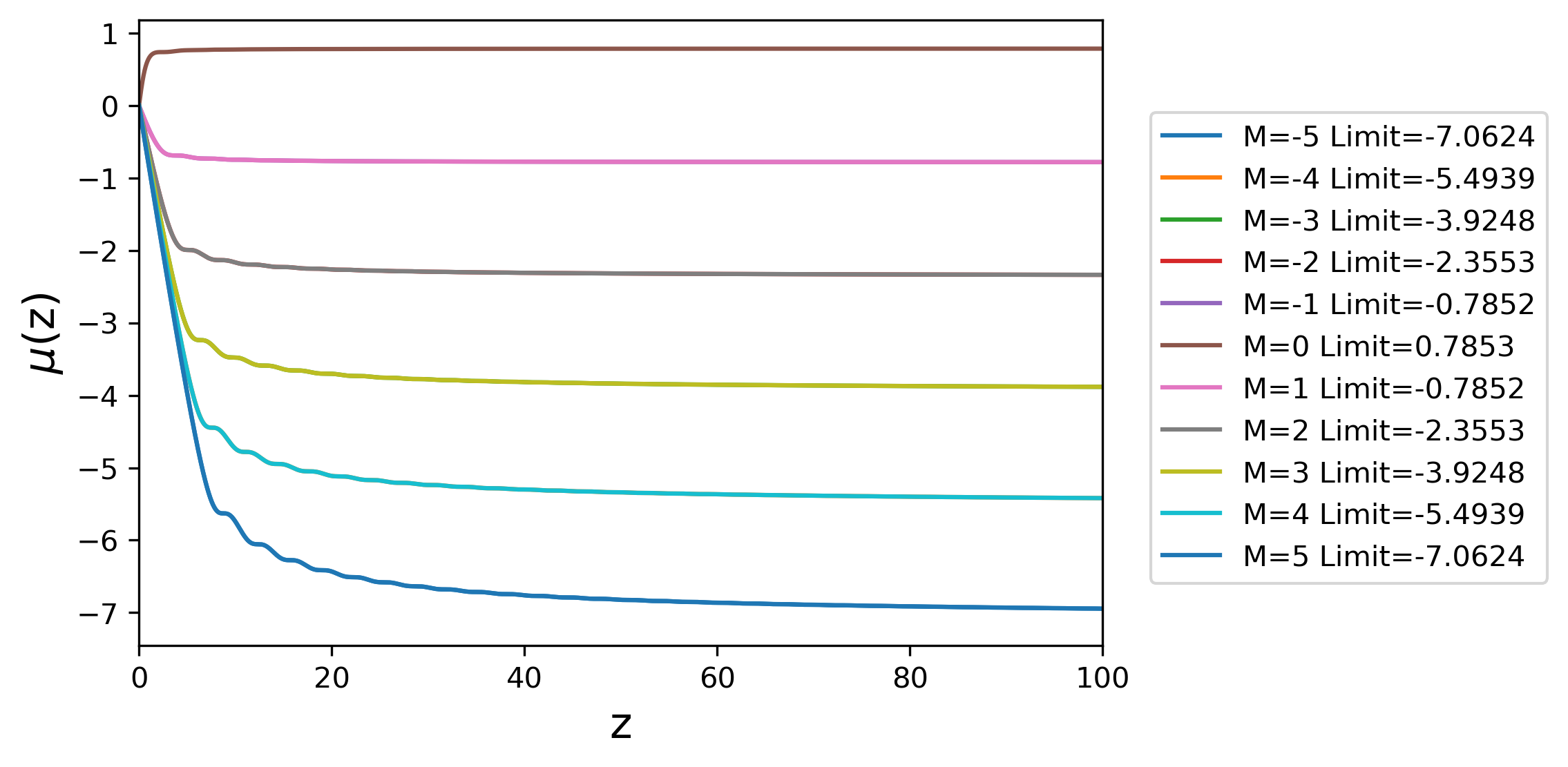}
	\caption{\(\mu(z)\) as a function of $z$ for $\lambda = 0$ and several values of $M$. The behaviour of positive and negative \(M\) are identical for all \(z\), since the governing equation depends only on $M^2$ in the $\lambda=0$ case.}
	\label{Fig plot of mu vs z}
\end{figure}

Although (\ref{Eq delta l phase shift}) cannot be evaluated exactly, it can be approximated very closely by choosing a sufficiently high value of \(z\). For \(\lambda = 0\), we have \(\mu(z\to\infty) = -l\pi/2 = (1 -2|M|)\pi/4\)  exactly, which allows us to check the accuracy of
the numerical computation of the phase shift when solving (\ref{Eq diff mu}) from $z=0$ up to a large enough $z$. This is illustrated in Table~\ref{table mu} and Fig.~\ref{Fig plot of mu vs z}. The results for negative and positive values of $M$ are the same, since when $\lambda=0$ the right-hand side of (\ref{Eq diff mu}) depends only on $M^2$.

To solve (\ref{Eq diff mu}) using the Runge-Kutta method the initial conditions need to be specified. At \(z=0\) we must have a solution equal to zero, so from (\ref{Eq phase shift solution z}), \(\mu(0)=0\). To remove the singularity at $z=0$ in (\ref{Eq diff mu}), we have
\begin{equation}
	\centering
	\mu(z)=\mu_0z + \mathcal{O}(z^2).
\end{equation}
Then in the limit \(z\rightarrow0\) (\ref{Eq diff mu}) becomes
\begin{equation}
	\mu_0=-\left(M^2-\frac{1}{4}\right)\left(\mu_0+1\right)^2.
    \label{eq appendix mu 0}
\end{equation} 
This has two solutions for $\mu_0$, but we must choose the one that leads to the correct asymptotic behavior $u(z) \sim z^{|M|+1/2}$ at 
$z\to 0$. Therefore, we have
\begin{equation}
	\centering
	\frac{d \mu}{d z}(0) =\mu_0=\begin{cases}
		\displaystyle -\frac{M-\frac{1}{2}}{M+\frac{1}{2}} & \qquad M\geq0; \\
        & \\
		\displaystyle -\frac{M+\frac{1}{2}}{M - \frac{1}{2}} & \qquad M\leq0.	
	\end{cases}
\end{equation}

\begin{table}
	\centering
    	\begin{tabular}{|c|c|c|c|c|c|}
		\hline 
		$z$  & \(M=-4\) & \(M=-3\) & \(M=-2\) & \(M=-1\) & \(M=0\) \\
		\hline
		100  & $-5.41878629$ & $-3.88335873$ & $-2.33739599$ & $-0.78165724$ & $0.78414519$ \\
		\hline
		1000 & $-5.48991364$ & $-3.92261499$ & $-2.35431984$ & $-0.78502309$ & $0.78527319$\\ 
		\hline
		2000 & $-5.49385036$ & $-3.92480292$ & $-2.35525716$ & $-0.78521063$ & $0.78533568$ \\
		\hline
		\(\infty\) & $-5.49778714$ & $-3.92699081$ & $-2.35619449$ & $-0.78539816$ & $0.78539816$ \\
		\hline
	\end{tabular}
	\begin{tabular}{|c|c|c|c|c|c|}
		\hline 
		$z$  & \(M=1\) & \(M=2\) & \(M=3\) & \(M=4\)  \\
		\hline
		100  & $-0.78165724$ & $-2.33739599$ & $-3.88335873$ & $-5.41878629$ \\
		\hline
		1000 & $-0.78502309$ & $-2.35431984$ & $-3.92261499$ & $-5.48991364$ \\
		\hline 
		2000 & $-0.78521063$ & $-2.35525716$ & $-3.92480292$ & $-5.49385036$ \\
		\hline
		\(\infty\) & $-0.78539816$ & $-2.35619449$ & $-3.92699081$ & $-5.49778714$ \\
		\hline
	\end{tabular}
\caption{Tables comparing the values \(\mu(z)\) for various $M$ when \(\lambda=0\). The value of \(\mu(2000)\) is >99.9\% of the value of \(\mu(\infty)\) for all values of \(M\). }
\label{table mu}
\end{table}
.

%#################bibliography########################
\newpage
\bibliographystyle{RS}
\bibliography{Reference.bib}

\begin{thebibliography}{99}

\bibitem{RefWorks:33}
Dirac P. 1931  Quantised singularities in the electromagnetic field. {\em
  Proceedings of the Royal Society of London. Series A, Containing Papers of a
  Mathematical and Physical Character} \textbf{133}, 60--72.
(\href{http://dx.doi.org/10.1098/rspa.1931.0130}{10.1098/rspa.1931.0130})

\bibitem{Castelnovo2008}
Castelnovo C, Moessner R, Sondhi SL. 2008  Magnetic monopoles in spin ice. {\em
  Nature} \textbf{451}, 42--45.
(\href{http://dx.doi.org/10.1038/nature06433}{10.1038/nature06433})

\bibitem{Bramwell2020}
Bramwell ST, Harris MJ. 2020  The history of spin ice. {\em Journal of Physics:
  Condensed Matter} \textbf{32}, 374010.
(\href{http://dx.doi.org/10.1088/1361-648X/ab8423}{10.1088/1361-648X/ab8423})

\bibitem{Moller2006}
Möller G, Moessner R. 2006  Artificial Square Ice and Related Dipolar
  Nanoarrays. {\em Physical Review Letters} \textbf{96}, 237202.
(\href{http://dx.doi.org/10.1103/PhysRevLett.96.237202}{10.1103/PhysRevLett.96.237202})

\bibitem{Ladak2010}
Ladak S, Read DE, Perkins GK, Cohen LF, Branford WR. 2010  Direct observation
  of magnetic monopole defects in an artificial spin-ice system. {\em Nature
  Physics} \textbf{6}, 359--363.
(\href{http://dx.doi.org/10.1038/nphys1628}{10.1038/nphys1628})

\bibitem{Chern2011}
Chern GW, Mellado P, Tchernyshyov O. 2011  Two-Stage Ordering of Spins in
  Dipolar Spin Ice on the Kagome Lattice. {\em Physical review letters}
  \textbf{106}, 207202.
(\href{http://dx.doi.org/10.1103/PhysRevLett.106.207202}{10.1103/PhysRevLett.106.207202})

\bibitem{Wang2006}
Wang RF, Nisoli C, Freitas RS, Li J, McConville W, Cooley BJ, Lund MS, Samarth
  N, Leighton C, Crespi VH, Schiffer P. 2006  Artificial ‘spin ice’ in a
  geometrically frustrated lattice of nanoscale ferromagnetic islands. {\em
  Nature} \textbf{439}, 303--306.
(\href{http://dx.doi.org/10.1038/nature04447}{10.1038/nature04447})

\bibitem{RefWorks:17}
B{\'e}ch{\'e} A, Van~Boxem R, Van~Tendeloo G, Verbeeck J. 2014  Magnetic
  monopole field exposed by electrons. {\em Nature Physics} \textbf{10},
  26--29.
(\href{http://dx.doi.org/10.1038/nphys2816}{10.1038/nphys2816})

\bibitem{Khomskii2014}
Khomskii DI. 2014  Magnetic monopoles and unusual dynamics of magnetoelectrics.
  {\em Nature Communications} \textbf{5}, 4793.
(\href{http://dx.doi.org/10.1038/ncomms5793}{10.1038/ncomms5793})

\bibitem{Fechner2014}
Fechner M, Spaldin NA, Dzyaloshinskii IE. 2014  Magnetic field generated by a
  charge in a uniaxial magnetoelectric material. {\em Phys. Rev. B}
  \textbf{89}, 184415.
(\href{http://dx.doi.org/10.1103/PhysRevB.89.184415}{10.1103/PhysRevB.89.184415})

\bibitem{Meier2019}
Meier QN, Fechner M, Nozaki T, Sahashi M, Salman Z, Prokscha T, Suter A,
  Schoenherr P, Lilienblum M, Borisov P, Dzyaloshinskii IE, Fiebig M, Luetkens
  H, Spaldin NA. 2019  Search for the Magnetic Monopole at a Magnetoelectric
  Surface. {\em Phys. Rev. X} \textbf{9}, 011011.
(\href{http://dx.doi.org/10.1103/PhysRevX.9.011011}{10.1103/PhysRevX.9.011011})

\bibitem{Trugenberger2020}
Trugenberger C, Diamantini MC, Poccia N, Nogueira FS, Vinokur VM. 2020
  Magnetic Monopoles and Superinsulation in Josephson Junction Arrays. {\em
  Quantum Reports} \textbf{2}, 388.
(\href{http://dx.doi.org/10.3390/quantum2030027}{10.3390/quantum2030027})

\bibitem{Diamantini2021}
Diamantini MC, Trugenberger CA, Vinokur VM. 2021  Quantum magnetic monopole
  condensate. {\em Communications Physics} \textbf{4}, 25.
(\href{http://dx.doi.org/10.1038/s42005-021-00531-5}{10.1038/s42005-021-00531-5})

\bibitem{Birkeland1896}
Birkeland K. 1896  The Cathode Rays under the influence of strong magnetic
  forces. {\em Elec. Rev.} \textbf{38}, 752--782.

\bibitem{Poincare1896}
Poincaré H. 1896  Remarques sur une expérience de M. Birkeland. {\em Comptes
  Rendus de l'Académie des sciences} \textbf{123}, 530--533.

\bibitem{Ando1975}
Ando T, Matsumoto Y, Uemura Y. 1975  Theory of Hall Effect in a Two-Dimensional
  Electron System. {\em Journal of the Physical Society of Japan} \textbf{39},
  279--288.
(\href{http://dx.doi.org/10.1143/JPSJ.39.279}{10.1143/JPSJ.39.279})

\bibitem{Leblanc2012}
Leblanc L, Jiménez-García K, Williams R, Beeler M, Perry A, Phillips W,
  Spielman I. 2012  Observation of a superfluid Hall effect. {\em Proceedings
  of the National Academy of Sciences of the United States of America}
  \textbf{109}, 10811--10814.
(\href{http://dx.doi.org/10.1073/pnas.1202579109}{10.1073/pnas.1202579109})

\bibitem{Diamantini2022}
Diamantini MC, Trugenberger CA, Vinokur VM. 2021  Effective magnetic monopole
  mechanism for localized electron pairing in HTS. {\em Frontiers in Physics}
  \textbf{10}, 909310.
(\href{http://dx.doi.org/10.3389/fphy.2022.909310}{10.3389/fphy.2022.909310})

\bibitem{RefWorks:36}
Sakurai JJ. 1994 {\em Modern Quantum Mechanics}.
Menlo Park, Calif. [u.a.]: Benjamin/Cummins 1st edition.

\bibitem{RefWorks:37}
Griffiths DJ. 1981 {\em Introduction to Electrodynamics}.
Cambridge: Cambridge University Press 4th edition.

\bibitem{RefWorks:1}
Landau DL, Lifshitz EM. 1977 pp. 164--198.
In {\em The Quasi-Classical Case}, Quantum Mechanics (Non-relativistic Theory)
  Volume 3 pp. 164--198. Linacre House, Jordan Hill, Oxford, OX2 8DP:
  Butterworth-Heinemann 3rd edition.

\bibitem{RefWorks:29}
Kittel C. 1987 {\em Quantum Theory of Solids}.
Chichester: John Wiley and Sons 2nd edition.

\bibitem{RefWorks:34}
Greiner W. 1998 {\em Quantum Mechanics an Introduction}.
Berlin: Springer 3rd edition.

\bibitem{RefWorks:35}
Griffiths DJ, Schroeter DF. 2013 {\em Introduction to Quantum Mechanics}.
Harlow: Pearson 3rd edition.

\bibitem{RefWorks:10}
Argyres PN. 1965  The Bohr-Sommerfeld quantization rule and the Weyl
  correspondence. {\em Physics Physique Fizika} \textbf{2}, 131--139.
(\href{http://dx.doi.org/10.1103/PhysicsPhysiqueFizika.2.131}{10.1103/PhysicsPhysiqueFizika.2.131})

\bibitem{RefWorks:11}
Tarnovskii AS. 1990  The Bohr-Sommerfeld quantization rule and quantum
  mechanics. {\em Soviet Physics Uspekhi} \textbf{33}, 86--86.
(\href{http://dx.doi.org/10.1070/PU1990v033n01ABEH002407}{10.1070/PU1990v033n01ABEH002407})

\bibitem{RefWorks:14}
Cushman R, Śniatycki J. 2014  {On Bohr-Sommerfeld-Heisenberg Quantization}.
  {\em Journal of Geometry and Symmetry in Physics} \textbf{35}, 11 -- 19.
(\href{http://dx.doi.org/10.7546/jgsp-35-2014-11-19}{10.7546/jgsp-35-2014-11-19})

\bibitem{RefWorks:16}
Ishkhanyan AM, Krainov VP. 2017  Maslov index for power-law potentials. {\em
  JETP Letters} \textbf{105}, 43--46.
(\href{http://dx.doi.org/10.1134/S0021364017010106}{10.1134/S0021364017010106})

\bibitem{RefWorks:5}
Robbin J, Salamon D. 1993  The Maslov index for paths. {\em Topology}
  \textbf{32}, 827--844.
(\href{http://dx.doi.org/10.1016/0040-9383(93)90052-W}{10.1016/0040-9383(93)90052-W})

\bibitem{RefWorks:20}
Keller JB. 1958  Corrected bohr-sommerfeld quantum conditions for nonseparable
  systems. {\em Annals of Physics} \textbf{4}, 180--188.
(\href{http://dx.doi.org/10.1016/0003-4916(58)90032-0}{10.1016/0003-4916(58)90032-0})

\bibitem{RefWorks:24}
M\"uller-Kirsten HJW. 2012 {\em Introduction to quantum mechanics}.
Singapore: World Scientific 1st edition.

\bibitem{RefWorks:21}
Merzbacher E. 1998 {\em Quantum Mechanics}.
New York: John Wiley and Sons 2nd edition.

\bibitem{RefWorks:22}
Jammer M. 1989 {\em The conceptual development of quantum mechanics}.
Los Angeles: Tomash 1 edition.

\bibitem{RefWorks:23}
Weinberg S. 2018 {\em Lectures on quantum mechanics}.
Cambridge: Cambridge University Press 2nd edition.

\bibitem{RefWorks:4}
Eisberg R, Resnick R. 1985 pp. 110--116.
In {\em Interpretation of the Quantization Rules, Sommefields Model}, Quantum
  Physics of Atoms, Molecules, Solids, Nuclei, and Particles pp. 110--116.
  Canada: John Wiley and Sons 2nd edition.

\bibitem{Paz2000}
Paz G. 2000a  The non-self-adjointness of the radial momentum operator in n
  dimensions. {\em Journal of Physics A: Mathematical and General} \textbf{35}.
(\href{http://dx.doi.org/10.1088/0305-4470/35/16/311}{10.1088/0305-4470/35/16/311})

\bibitem{Paz2000(1)}
Paz G. 2000b  On the connection between the radial momentum operator and the
  Hamiltonian in n dimensions. {\em European Journal of Physics} \textbf{22}.
(\href{http://dx.doi.org/10.1088/0143-0807/22/4/308}{10.1088/0143-0807/22/4/308})

\bibitem{RefWorks:2}
Bransden BH, Joachain CJ. 2000 pp. 408--419, 783--785.
In {\em 8 Approximation Methods for Stationary Systems, 8.4 The WKB
  approximation}, Quantum Mechanics pp. 408--419, 783--785. Edinburgh Gate,
  Harlow, Essex, CM20 2JE: Pearson Education Ltd 2nd edition.

\bibitem{RefWorks:3}
Simmonds JG, Jr JEM. 1998 pp. 71--80.
In {\em The WKB Approximation}, A First Look at Perturbation Theory pp. 71--80.
  Mineola, NewYork: Dover Publications , INC 2nd edition.

\bibitem{RefWorks:18}
Morse MP, Allis WP. 1933  The Effect of Exchange on the Scattering of Slow
  Electrons from Atoms. {\em Phys. Rev.} \textbf{44}, 269--276.
(\href{http://dx.doi.org/10.1103/PhysRev.44.269}{10.1103/PhysRev.44.269})

\bibitem{Qi2009}
Qi XL, Li R, Zang J, Zhang SC. 2009  Inducing a Magnetic Monopole with
  Topological Surface States. {\em Science} \textbf{323}, 1184--1187.
(\href{http://dx.doi.org/10.1126/science.1167747}{10.1126/science.1167747})

\bibitem{Uri2020}
Uri A, Kim Y, Bagani K, Lewandowski CK, Grover S, Auerbach N, Lachman EO,
  Myasoedov Y, Taniguchi T, Watanabe K, Smet J, Zeldov E. 2020  Nanoscale
  imaging of equilibrium quantum Hall edge currents and of the magnetic
  monopole response in graphene. {\em Nature Physics} \textbf{16}, 164--170.
(\href{http://dx.doi.org/10.1038/s41567-019-0713-3}{10.1038/s41567-019-0713-3})

\bibitem{Diamantini2025}
Diamantini MC. 2025  Emergent Magnetic Monopoles in Quantum Matter. {\em
  Condensed Matter} \textbf{10}.
(\href{http://dx.doi.org/10.3390/condmat10020020}{10.3390/condmat10020020})

\bibitem{RefWorks:32}
Ryder LH. 1996 {\em Quantum Field Theory}.
Cambridge: Cambridge University Press 2nd edition.

\end{thebibliography}

\end{document}